\documentclass[twocolumn]{aastex631}

\usepackage{savesym}
\savesymbol{tablenum}
\usepackage{siunitx}
\DeclareSIUnit\Msun{\ensuremath{\mathrm{M}_{\odot}}}
\DeclareSIUnit\erg{\ensuremath{\mathrm{erg}}}
\DeclareSIUnit\angstrom{\text {Å}}
\restoresymbol{SIX}{tablenum}

\usepackage[version=3]{mhchem}

\graphicspath{{./}{figures/}}

\hypersetup{linkcolor=red,citecolor=magenta,urlcolor=blue}

\newcommand{\tuc}{47\ Tucanae}
\newcommand{\omegacen}{$\omega$\,Centauri}
\newcommand{\teff}{T_\mathrm{eff}}
\newcommand{\logg}{\log_{10}(g)}
\newcommand{\review}[1]{#1}

\begin{document}

\title{Exploring the Chemistry and Mass Function of the Globular Cluster \tuc{} with New Theoretical Color-Magnitude Diagrams}

\author[0000-0003-0398-639X]{Roman Gerasimov}
\affiliation{Center for Astrophysics and Space Sciences,
University of California, San Diego, La Jolla, California 92093, USA
}
\affiliation{Department of Physics and Astronomy, University of Notre Dame, Nieuwland Science Hall, Notre Dame, IN 46556, USA}

\author[0000-0002-6523-9536]{Adam J.\ Burgasser}
\affiliation{Center for Astrophysics and Space Sciences,
University of California, San Diego, La Jolla, California 92093, USA
}

\author[0000-0002-4770-5388]{Ilaria Caiazzo}
\affiliation{Division of Physics, Mathematics and Astronomy, California Institute of Technology, Pasadena, CA 91125, USA}

\author[0000-0002-8546-9128]{Derek Homeier}
\affiliation{Aperio Software Ltd., Insight House, Riverside Business Park, Stoney Common Road,
Stansted, Essex, CM24 8PL, United Kingdom}

\author[0000-0001-9002-8178]{Harvey B.\ Richer}
\affiliation{Department of Physics and Astronomy, University of British Columbia, 6224 Agricultural Road, Vancouver, BC V6T 1Z1, Canada}

\author[0000-0001-6464-3257]{Matteo Correnti}
\affiliation{INAF Osservatorio Astronomico di Roma, Via Frascati 33, 00078, Monteporzio Catone, Rome, Italy}
\affiliation{ASI-Space Science Data Center, Via del Politecnico, I-00133, Rome, Italy}

\author[0000-0001-9002-8178]{Jeremy Heyl}
\affiliation{Department of Physics and Astronomy, University of British Columbia, 6224 Agricultural Road, Vancouver, BC V6T 1Z1, Canada}

\begin{abstract}

Despite their shared origin, members of globular clusters display star-to-star variations in composition. The observed pattern of element abundances is unique to these stellar environments, and cannot be fully explained by any proposed mechanism. It remains unclear whether stars form with chemical heterogeneity, or inherit it from interactions with other members. These scenarios may be differentiated by the dependence of chemical spread on stellar mass; however, obtaining a sufficiently large mass baseline requires abundance measurements on the lower main sequence that is too faint for spectroscopy even in the nearest globular clusters. We developed a stellar modelling method to obtain precise chemical abundances for stars near the end of the main sequence from multiband photometry, and applied it to the globular cluster \tuc{}. The computational efficiency is attained by matching chemical elements to the model components that are most sensitive to their abundance. We determined \ce{[O/Fe]} for  $\sim5000$ members below the main sequence \review{knee} at the level of accuracy, comparable to the spectroscopic measurements of evolved members in literature. The inferred distribution disfavors stellar interactions as the origin of chemical spread; however, an accurate theory of accretion is required to draw a more definitive conclusion. We anticipate that future observations of \tuc{} with \textit{JWST} will extend the mass baseline of our analysis into the substellar regime. Therefore, we present predicted color-magnitude diagrams and mass-magnitude relations for the brown dwarf members of \tuc{}.

\end{abstract}


\section{Introduction} \label{sec:introduction}
Globular clusters are large gravitationally bound conglomerations of stars, commonly observed in the stellar halos of galaxies including our own \citep{Harris_review_1,Harris_review_2,GCs_in_M31}. Since the brightest examples (e.g. \omegacen{}, \tuc{}) are prominent naked-eye targets, the study of globular clusters dates back multiple centuries \citep{lacaille_catalog,messier}. A typical globular cluster hosts over $\num{e5}$ members \citep{GC_masses}, far outnumbering the populations of most open clusters ($\sim20$ -- $\num{e4}$ members, \citealt{OC_masses,OC_masses_2}) that form in the present-day Milky Way. This points to the origin of globular clusters in the early phases of the hierarchical assembly of galaxies \citep{milky_way_assembly}, making them some of the oldest objects in the universe, often exceeding $\qty{12}{Gyr}$ in age \citep{GC_ages,GC_ages_2}. %In fact, the oldest globular clusters could have originated in the first star-forming structures to emerge after the Big Bang \citep{pre_reionization_GC,pre_reionization_GC_3,pre_reionization_GC_4,pre_reionization_GC_2}, and may predate the onset of galactic formation \citep{pregalactic_GCs,pregalactic_GCs_2}.
By consequence, the members of most globular clusters exhibit sub-solar metallicities ($\ce{[Fe/H]}\lesssim\qty{-1}{dex}$, \citealt{GC_metallicities,GC_distances,GC_first_metal_poor}) and have a distinct color distribution \citep{GC_CMD_difference_1,GC_CMD_difference_2,GC_CMD_HB} that motivated the original definition of stellar populations \citep{baade_populations,Oort_populations}.

Globular clusters are prime stellar astrophysics laboratories \citep{GCs_as_laboratories_2} and probes of galactic formation and evolution \citep{GC_archaeology,GC_archaeology_2,GC_archaeology_3,GC_archaeology_4}. The old ages of globular clusters provide a direct constraint on the age of the universe \citep{GC_age_universe_3,GC_age_universe,GC_age_universe_2}, while their kinematic properties trace the distribution of dark matter \citep{GC_DM_1,GC_DM_2,GC_DM_3}. Furthermore, tidally disrupted globular clusters are responsible for the production of stellar streams, alongside dwarf galaxies \citep{streams_review}. Understanding the properties and evolution of these unique objects is therefore of uttermost importance to stellar astrophysics, galactic science and cosmology.

Naively, the coeval nature of star clusters leads to the expectation of uniformity in element abundances among their members. Nonetheless, the presence of star-to-star chemical variations within globular clusters was evident since the early spectroscopic observations that found differing cyanogen (\ce{CN}) absorption strength \citep{early_CN_2} in stars of \review{similar} spectral types and luminosity classes (the ``\textit{cyanogen discrepancy}'', \citealt{cyanogen_discrepancy}), followed by the first discoveries of carbon star members \citep{first_C_star,second_C_star}. At first, these anomalies were largely attributed to evolutionary effects rather than a genuine primordial inhomogeneity (e.g., \citealt{early_evo_effects_1,early_evo_effects_2,early_evo_effects_3,meridional_circulation,early_evo_effects_4}), in part, because the comparable features of peculiar field stars were generally consistent with internal processing \citep{CH_stars,CH_stars_2}. This consensus was eventually challenged by the observed spread in \ce{CNO} abundances at early evolutionary stages \citep{evolution_failure_1,evolution_failure_2,no_GC_abundances_in_field}, evidence of variations in \ce{[Na/Fe]} and \ce{[Al/Fe]} \citep{early_NaAl_3,NaO_anticorrelation,early_NaAl_1,early_NaAl_2} and the discovery of heavy element (atomic number greater than $13$) variations in the globular cluster \omegacen{} \citep{omega_cen_calcium}.

Closely related to chemical heterogeneity is the so-called \textit{second parameter problem}, characterized by the diversity of horizontal branch morphologies among color-magnitude diagrams (CMDs) of globular clusters with comparable metallicities \citep{second_parameter_1,second_parameter_2}. In some cases, the horizontal branch is distinctly bimodal (e.g., NGC 2808, \citealt{HB_gap}), which led \citet{multiple_populations} (also see \citealt{multiple_populations_2}, \review{as well as earlier speculation by \citealt{early_NaAl_1,early_NaAl_2}}) to suggest the presence of \textit{multiple populations} (MP) of stars with distinct histories of chemical enrichment. In this scenario, a fraction of members (the \textit{primordial population}, using the terminology of \citealt{review_main}) form similarly to regular halo stars, while the rest (the \textit{enriched population}) are influenced by the unique environment of the cluster. \review{The connection between the horizontal branch morphology and chemical anomalies was first investigated by \citet{early_evo_effects_4,early_NaAl_2,bimodality_correlation_2,bimodality_correlation_3,bimodality_correlation}} on the basis of the observed bimodality in \ce{CN} absorption among globular clusters with split horizontal branches. This hypothesis is now firmly established based on detailed spectroscopic abundance measurements of the horizontal branch members \citep{second_parameter_solution}.

Over the last three decades, the existence of MP in globular clusters has been confirmed by detailed spectroscopic \review{analysis} of giant (e.g., \citealt{CNO_sum_constant,spectroscopy_giants_1,nominal_T14,nominal_C14,spectroscopy_giants_2,spectroscopy_giants_3}), subgiant and \review{upper main sequence} stars (e.g., \citealt{nominal_M16,spectroscopy_MSTO_1,spectroscopy_MSTO_2}), integrated spectroscopy \citep{47_Tuc_age_review}, and splitting of main sequence (e.g., \citealt{anderson_thesis,bedin_bifurcation,MS_splitting_2,MS_splitting_1}) and post-main sequence (e.g., \citealt{postMS_splitting_1,postMS_splitting_2}) CMDs. A more comprehensive summary of available evidence of MP may be found in numerous reviews of the subject (\citealt{review_main,review_2,review_3,review_4,review_5,review_6,second_parameter_dotter} and references within).

The enriched population is characterized by distinct light element abundances: most prominently, the overabundance of nitrogen and sodium, and the depletion of carbon and oxygen, compared to the primordial population. On the other hand, the abundances of heavy elements generally do not display star-to-star variations (globular clusters are said to be \textit{mono-metallic}, in the sense that $\ce{[Fe/H]}\approx\mathrm{const}$), with the exception of a few \textit{anomalous globular clusters} \citep{anomalous_GCs}, of which \omegacen{} is the most well-known \citep{omegacen_anomaly_1,omegacen_anomaly_2,oMEGACat}. %An even smaller number of globular clusters also display large dispersions of $r$-process elements (e.g. barium, europium; \citealt{evan_1}, 2023 in preparation).
By contrast, these characteristic abundance patterns are not observed in open clusters \citep{oc_evolution,oc_evolution_2,oc_evolution_3} and are rare ($\sim3\%$ of the $[\ce{Fe/H}]\leq-1$ local sample, \citealt{GC_ejected_ratio}) among field stars \citep{field_evolution_4,field_evolution,field_evolution_2,field_evolution_3}, motivating the common assumption that the handful of known examples were ejected from globular clusters in the past \citep{GC_ejected_stars,GC_ejected_stars_2}.

The physical origin of MP remains largely unexplained \citep{review_main}. The mono-metallic nature of non-anomalous globular clusters is usually attributed to the high velocity of supernova ejecta that expel a significant amount of pristine gas from the cluster, alongside most of the material enriched with heavy elements, \review{within $\sim10-35\ \qty{}{Myr}$ of the onset of star formation \citep{AGB_SN_modulation_2,AGB_SN_modulation}}. Many of the proposed MP theories fall under one of two broad categories. In \textit{multiple generation models} (e.g., \citealt{early_NaAl_1,FRMS,SMS}), star formation proceeds in two or more bursts with a sufficient time interval to allow the earlier generations of stars to build up a reservoir of enriched gas, from which later generations are assembled. Alternatively, \textit{concurrent formation models} (e.g., \citealt{early_disk_accretion,concurrent_SMS,BD_accretion}) invoke a single generation of coeval stars, in which a fraction of the members are polluted by stellar ejecta to attain the chemical abundances of the enriched population. Both approaches suffer from major shortcomings. Since nuclear processing primarily occurs in massive stars, and since these stars make a subdominant contribution to the overall mass budget for commonly assumed initial mass functions (e.g., \citealt{Kroupa}), it is unclear how a sufficient amount of processed material can be produced in multiple generation models to assemble the enriched population of stars (enriched stars make up the majority of members in most globular clusters, \citealt{population_ratios}). This \textit{mass budget problem} is easier to resolve in concurrent formation models, where the enriched population is envisioned to form from pristine gas prior to the onset of supernovae; however, considerable fine-tuning is needed to match the required pollution timescales \citep{disk_accretion_timescales} and to reproduce the discreetness of populations that arises naturally when multiple generations are involved.

A potential observational signature of concurrent formation models is the possible dependence of element abundances on the initial stellar mass \citep{JWST_low_MS_phot_1,JWST_low_MS_phot_2}, related to the fact that the magnitude of chemical enrichment is determined not only by the composition of the polluted material produced by the primordial generation, but also by the ability of the enriched generation of stars to accrete it. If the polluted material is primarily accreted onto circumstellar disks, the accretion efficiency is expected to be proportional to the surface area of the disk and its longevity, which, in turn, depends on the mass and evolution of the parent star \citep{disk_mass_dependence_1,disk_mass_dependence_2}. If instead the material is accreted onto the surface of the star, the accretion rate is expected to be proportional to the squared stellar mass (e.g., in the Bondi accretion formalism, \citealt{Bondi}). The accreted material may then be diluted by convective mixing within the star, which also depends on the stellar mass \citep{BC_origin}. Finally, the density of polluted material may not be uniform throughout the cluster. For example, in the concurrent formation model of \citet{early_disk_accretion}, the polluted material is only available in the core of the cluster. In this case, the accretion efficiency will depend on the kinematic properties of the stars, which are also related to stellar mass \citep{omega_cen_equipartition_coefficient,omega_cen_equipartition_segregation}.

Due to the large distances to globular clusters ($\gtrsim\qty{5}{kpc}$, \citealt{GC_distances}), high signal-to-noise ratio spectra can only be obtained for the brightest members that fall within a narrow mass range. Studying the variation of chemical abundances over a wider mass range inevitably requires the inference of chemistry from multiband photometry. In general, the photometric colors of low-mass stars are highly sensitive to the variations in element abundances due to the dominant molecular chemistry and opacity in low-temperature atmospheres \citep{clouds_and_chemistry,coolstars_21}. Furthermore, low-mass stars are particularly valuable as chemical tracers, since molecular absorption is less affected by the non-equilibrium radiation field \citep{NLTE_molecules}, while the interiors of low-mass stars are fully mixed and undergo minimal nuclear processing \citep{BC_origin,roman_omega_cen}. Sophisticated stellar models are required to take full advantage of this potential. Accurate simulation of the intricate physics and chemistry in low-temperature atmospheres remains extremely challenging.% Of particular importance is the effect of chemical abundances on stellar evolution \citep{BC_origin,MIST,ATMO_boundary_conditions,roman_omega_cen} that becomes prominent at the effective temperatures of $\teff{}\lesssim\qty{4500}{\K}$.

Considerable progress in the photometric analysis of cool stars near the end of the main sequence ($\gtrsim\qty{0.1}{\Msun}$) in globular clusters has been made since the advent of space-based observations with the \textit{Hubble Space Telescope} (\textit{HST}) and the \textit{James Webb Space Telescope} (\textit{JWST}). Recent results include measurement of the \ce{[O/Fe]} spread in the globular clusters NGC 6752 \citep{NGC_6752_spread}, M92 \citep{JWST_low_MS_phot_1} and \tuc{} \citep{JWST_low_MS_phot_2}, photometric characterization of individual populations in NGC 6752 \citep{NGC_6752_individual,coolstars_21}, and measurement of $\alpha$-enhancement in \omegacen{} \citep{roman_omega_cen}.

The CMDs of globular clusters are expected to extend far beyond the end of the main sequence into the brown dwarf regime. Unlike stars, brown dwarfs do not establish energy equilibrium and, instead, undergo long-term cooling, thereby creating a stellar/substellar gap in the CMD between the faintest main sequence star and the brightest brown dwarf \citep{adam_gap,age_gap,GC_whitepaper}. Photometric observations of the substellar sequence in globular clusters would then not only extend the mass baseline of MP studies, but also provide an independent constraint on the cluster age from the width of the gap. The faint magnitudes of globular cluster brown dwarfs pose a major challenge to their detection. The results of dedicated searches for brown dwarf candidates in the globular cluster M4 \citep{BD_hunt_1,BD_hunt_2} with \textit{HST} remain inconclusive. In our previous work \citep{roman_omega_cen}, we estimated the colors and magnitudes of the brown dwarfs in the globular cluster \omegacen{} and concluded that the substellar sequences of nearby globular clusters fall within the sensitivity limit of \textit{JWST}. Since then, \citet{rolly_BD} have reported the first tentative discovery of a brown dwarf in the globular cluster \tuc{}, designated \textit{BD10}. While the substellar nature of this object remains to be confirmed, we anticipate that many similar discoveries will follow in the near future from the ongoing Cycle 1 \textit{JWST} campaigns \citep{JWST_proposal_1,JWST_proposal_2} and subsequent cycles.

In this work, we describe a new method to determine the chemical abundances and other fundamental parameters of globular clusters based on their CMDs from the subgiant branch to the substellar regime. Our approach involves the computation of new evolutionary models and model atmospheres with fully self-consistent chemical abundances. Model isochrones are calculated and fitted to the observed distribution of photometric colors in an iterative manner. The computational efficiency of the process is attained by identifying the components of the stellar models that are most sensitive to particular elements, and recalculating them only when the abundances of those elements are updated. We apply our method to the brightest mono-metallic (non-anomalous) globular cluster \tuc{}.

This paper is organized as follows. In Section~\ref{sec:nominal_chem} we describe our compilation of spectroscopic \tuc{} abundances in the literature. The archival \textit{HST} photometry that was used in this study is briefly described in Section~\ref{sec:data}. Section~\ref{sec:isochrones} details the process of calculating a theoretical isochrone for a given set of chemical abundances, and includes a thorough analysis of associated systematic errors. Our method of isochrone fitting to the observed CMD is presented and applied to \tuc{} in Section~\ref{sec:chemistry}. In Section~\ref{sec:analysis}, we derive the mass function of the cluster and predict the anticipated properties of its substellar members. Our results are discussed and the conclusions are drawn in Section~\ref{sec:conclusion}.

\section{Nominal chemistry} \label{sec:nominal_chem}
In this study, we adopt spectroscopically inferred chemical abundances of \tuc{} as a baseline for comparison against their photometric counterparts, as well as an initial guess in CMD fitting (Section~\ref{sec:chemistry}). We refer to this set of abundances as the \textit{nominal} composition of the cluster. The nominal abundances are compiled from three sources: \citet{nominal_T14}, \citet{nominal_C14} and \citet{nominal_M16}. The set is similar to that used in \citet{sky_47_tuc}. Here, we provide a more complete description of its derivation.

\citet{nominal_T14} list metallicity (\ce{[Fe/H]}) and the abundances of $26$ other elements, measured for $13$ stars at the tip of the red giant branch ($\qty{3900}{ \kelvin}\lesssim \teff \lesssim \qty{4300}{ \kelvin}$, $0.3\lesssim \logg \lesssim 1.5$) in \tuc{}. For \ce{[Fe/H]}, \ce{[Ti/Fe]} and \ce{[Sc/Fe]}, separate measurements are made for the lines of neutral and ionized species. Both estimates of \ce{[Fe/H]} are consistent and treated as independent in our analysis. Since titanium and scandium have much lower ionization potentials than iron \citep{ionpot}, the neutral lines of these elements may be affected by unaccounted ultraviolet overionization, particularly prominent at low effective temperatures and metallicities \citep{NLTE_Ti,NLTE_Sc,NLTE_overview}. As such, we discard the neutral line measurements of \ce{[Ti/Fe]} and \ce{[Sc/Fe]} in \citet{nominal_T14}. We further discard all measurements without uncertainties.

\citet{nominal_C14} obtained $181$ composition measurements for $164$ unique red giant and asymptotic giant members of \tuc{} ($\logg\lesssim 3$). The dataset includes \ce{[Fe/H]} and $9$ other elements. Of $164$ observed stars, $5$ have also been analyzed in \citet{nominal_T14}. Abundance measurements reported in both references are generally consistent within the published error bars, with the exception of \ce{[Al/Fe]}, for which the value in \citet{nominal_C14} exceeds that of \citet{nominal_T14} by $\sim 2$ and $\sim 2.5$ sigma for the stars 2618 and 3622 respectively (catalog numbers from \citealt{Lee_catalog}). The difference may be partially caused by the more detailed modelling of \ce{Al} lines in \citet{nominal_T14}, following \citet{NLTE_Al}.

The measurements in \citet{nominal_M16} extend our sample to the fainter red giant and sub-giant members with $3<\logg<4$. The dataset lists metallicities and abundances of $7$ individual elements for $74$ stars. For sodium, we adopt the values with non-local thermodynamic equilibrium corrections based on the curves-of-growth from \citet{NLTE_Na}. Following the authors' recommendation, we use uncertainties of $\qty{0.1}{dex}$, $\qty{0.1}{dex}$ and $\qty{0.15}{dex}$ in \ce{[Mg/Fe]}, \ce{[Al/Fe]} and \ce{[Si/Fe]} respectively, instead of the quoted measurement errors due to the effect of \ce{CN} molecular features on the analyzed lines of these elements around $\qty{0.8}{\micro\metre}$.

The final sample includes measurements from multiple sources for \ce{[Fe/H]}, \ce{[Al/Fe]}, \ce{[Ca/Fe]}, \ce{[Eu/Fe]}, \ce{[La/Fe]}, \ce{[Mg/Fe]}, \ce{[Na/Fe]}, \ce{[Ni/Fe]}, \ce{[O/Fe]}, \ce{[Si/Fe]} and \ce{[Ti/Fe]}. We carried out a Kolmogorov–Smirnov statistical test for each of these elements to determine whether the measurements from different sources in literature are consistent with a shared parent population. The null-hypothesis (measurements are consistent) was rejected with $99\%$ confidence for \ce{[Al/Fe]}, \ce{[Eu/Fe]}, \ce{[Mg/Fe]}, \ce{[Na/Fe]}, \ce{[Ni/Fe]} and \ce{[Si/Fe]}. Since the spectroscopic measurements used in this study span a vast range of post-main sequence evolutionary stages, the discrepancy may reflect a genuine alteration of surface abundances by the first dredge-up \citep{red_giant_abundances_2}, meridional circulation \citep{meridional_circulation}, thermohaline mixing \citep{thermohaline_mixing} or other mechanisms. However, the effect is normally most pronounced in the products of the \ce{CNO} cycle \citep{red_giant_abundances_1}, all of which have passed the consistency test in our compilation. Therefore, the observed discrepancies between different literature sources are likely systematic in nature, exemplified by the aforementioned mismatch in aluminum measurements between \citet{nominal_T14} and \citet{nominal_C14}.

All spectroscopic composition measurements from the three literature sources were combined into a single set of chemical abundances, available in Table~\ref{tab:nominal_chem} of Appendix~\ref{sec:nominal_chem_table}. We assumed that each abundance measurement ($x_i^{(X)}$ for the $i$-th measurement of element $X$ out of $N^{(X)}$ measurements in total) is drawn from a normal distribution with the standard deviation composed of two components added in quadrature: the physical variation in chemistry among cluster members ($s^{(X)}$, identical for all measurements of $X$) and the measurement error ($\sigma_i^{(X)}$). The physical variation of each element may then be estimated as in Equation~\ref{eq:abundance_spread}:

\begin{equation}
    \label{eq:abundance_spread}
    \left(s^{(X)}\right)^2=\frac{N^{(X)} \mathrm{Var}\left(x_i^{(X)} \right)}{N^{(X)}-1.5}-\left\langle \left(\sigma_i^{(X)}\right)^2 \right\rangle
\end{equation}

Since we are ultimately interested in the standard deviation of the physical spread, we evaluated the unbiased sample variances assuming $N^{(X)}-1.5$ degrees of freedom \citep{unbiased_std}. In cases where the average measurement error exceeds the observed scatter in the data, $s^{(X)}$ cannot be estimated, and the data may be considered consistent with lack of star-to-star variations in $X$. The mean abundance of each element and its uncertainty were calculated using the square reciprocal measurement errors as statistical weights with $s^{(X)}$ added to the measurement error in quadrature if available. Finally, we note that the estimates of $s^{(X)}$ for the six elements that did not pass the Kolmogorov–Smirnov test are likely biased by systematic errors that may have been unaccounted for in the quoted measurement uncertainties.

Since the helium abundance cannot be measured spectroscopically at low effective temperatures, the nominal helium mass fraction of \tuc{} was adopted as $Y=0.25$, following the isochrone fit of eclipsing binary members in \citet{nominal_Y}.

\section{Archival photometry} \label{sec:data}
Throughout this paper, all mentions of the optical bands \texttt{F606W} and \texttt{F814W} will implicitly refer to the wide field channel of the \textit{Advanced Camera for Surveys} (\textit{ACS}) on \textit{HST}. Likewise, all mentions of the near infrared bands \texttt{F110W} and \texttt{F160W} will refer to the near infrared channel of the \textit{Wide Field Camera 3} (\textit{WFC3}) on \textit{HST}. Finally, the infrared bands \texttt{F150W2} and \texttt{F322W2} will refer to the \textit{Near Infrared Camera} (\textit{NIRCam}) on \textit{JWST}.

Our analysis is based on archival \textit{HST} photometry (\textit{GO-11677}, PI: H. Richer) of \tuc{} presented in \citet{data}. The primary field was imaged in the \texttt{F606W} and \texttt{F814W} bands. The field spans $\sim\qty{5}{arcmin}$ and is located at $\qty{6.7}{arcmin}$ ($2.1$ half-light radii, \citealt{GC_distances}) from the center of the cluster. Parallel fields were imaged in the \texttt{F110W} and \texttt{F160W} bands, as well as additional bands of the ultraviolet and visible light channel that are not used in this study. The parallel fields span a $250$ degree arc, centered at the primary field with the radius of $\sim\qty{6}{arcmin}$ and facing away from the center of the cluster (see Figure~2 of \citealt{data}). This work is based on the observations of the primary field and one of the parallel fields with the largest cumulative exposure time (field 13, referred to as the ``stare'').

The archival photometry is contaminated by the members of the Small Magellanic Cloud (SMC), centered $\sim2^\circ$ southeast of \tuc{}. Since the SMC is more distant than \tuc{} by over an order of magnitude, the main sequences of the two objects are well-separated in the optical CMD and have a small overlap in the near infrared CMD (see Figure~\ref{fig:best_fit}). The overlap region is excluded from our analysis as detailed in Section~\ref{sec:analysis}. A more thorough reduction of the archival data with proper motion cleaning is deferred to a future study.

All magnitudes quoted in this paper are \texttt{VEGAMAG}.

% \rom{Need Harvey's et al. help here to (1) verify that the above is correct and (2) to describe the proper motion measurements since I don't think they come from the same paper. (I downloaded them from Google Drive and could not figure out the original source)}

\section{Model isochrones} \label{sec:isochrones}
We derive the properties of \tuc{} by comparing theoretical model isochrones to the observed CMD of the cluster.  Due to the presence of MP, an accurate model of the CMD requires multiple theoretical isochrones that capture the observed spread in photometric colors. We therefore aim to find three isochrones with distinct chemical compositions that approximately trace out the \textit{blue tail}, the \textit{red tail} and the \textit{ridgeline} of the CMD. We adopt the convention of identifying the blue and red tails of the distribution based on the \texttt{F110W} -- \texttt{F160W} near infrared color of the lower main sequence (the optical colors are, in fact, inverted compared to their near infrared counterparts).

We search for the desired isochrones through iterative perturbations of the nominal chemical composition. Both the final isochrones and the intermediate isochrones at each iteration are calculated using an improved version of the method, originally developed for our previous study of the globular cluster \omegacen{} \citep{roman_omega_cen}. In this section, we describe the general process of calculating a theoretical isochrone for a given chemical composition.  Our approach to isochrone fitting and the treatment of multiple populations in the cluster are presented in Section~\ref{sec:chemistry}.

\subsection{Evolutionary models}

Stellar evolution is simulated using the \texttt{MESA} (\texttt{M}odules for \texttt{E}xperiments in \texttt{S}tellar \texttt{A}strophysics) code \citep{MESA,MESA_2,MESA_3,MESA_4,MESA_5}, version \texttt{15140}. The evolution begins at the pre-main sequence (PMS) phase and is allowed to proceed either until the age of $\qty{13.5}{Gyr}$ or until the tip of the subgiant branch, whichever occurs sooner. For our purposes, we define the tip of the subgiant branch as the point, at which the innermost shell that satisfies $\varepsilon_\mathrm{nuc}>\qty{e3}{\erg\per\gram\per\second}$ begins to fall outside the central $10\%$ of the stellar mass ($\varepsilon_\mathrm{nuc}$ is the specific nuclear energy release rate). This criterion was empirically determined to be a reliable indicator of the hydrogen shell ignition that characterizes the onset of the red giant branch \citep{RGB_definition}. Evolving the model further into the red giant branch incurs a larger computational cost due to the rapid variation of surface parameters with age, and is beyond the scope of this study; however, see \citet{red_giant_abundances_2}.

The initial stellar masses are sampled on an adaptive grid that guarantees the difference in luminosities and effective temperatures between adjacent grid masses of $\left|\Delta \log_{10}{L}\right|<\qty{0.12}{dex}$ and $\left|\Delta \teff\right|<\qty{120}{\K}$, respectively, at the \textit{checkpoint} ages between $\qty{10}{Gyr}$ and $\qty{13.5}{Gyr}$ in steps of $\qty{0.5}{Gyr}$ (the typical average differences at $\qty{13.5}{Gyr}$ are $\langle\left|\Delta \log_{10}{L}\right|\rangle\approx\qty{0.07}{dex}$ and $\langle\left|\Delta \teff\right|\rangle\approx\qty{70}{\K}$). The lowest mass in the grid ($\sim \qty{0.04}{\Msun}-\qty{0.05}{\Msun}$) is chosen to attain an evolved $\teff{}$ of $\lesssim\qty{700}{\K}$ at $\qty{13.5}{Gyr}$, while the highest mass ($\sim\qty{0.9}{\Msun}$) is set by requiring that the star takes at least $\qty{10}{Gyr}$ to reach the tip of the subgiant branch.

\begin{figure}[ht!]
    \centering
    \includegraphics[width=1\columnwidth]{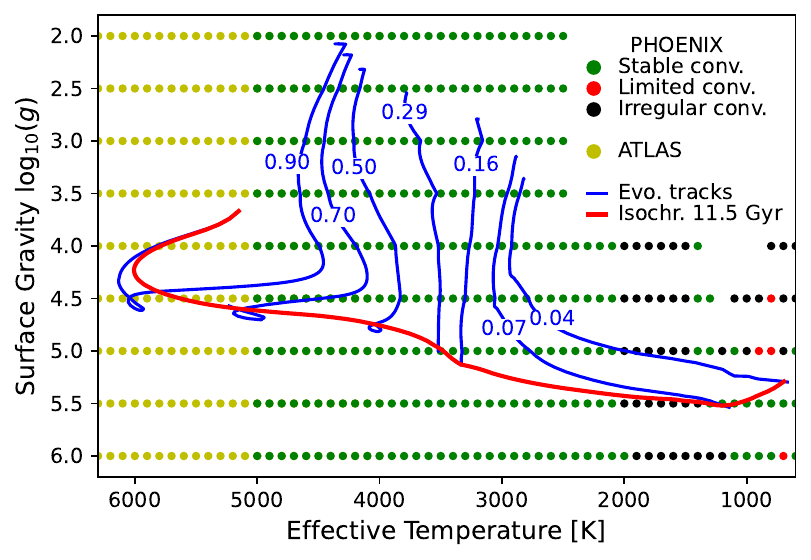}
    \caption{Effective temperatures and surface gravities (\review{Kiel diagram}) of the model atmospheres, calculated in this study for the final \textit{ridgeline} isochrone of \tuc{}. The \texttt{PHOENIX} models are color-coded by their convergence patterns, as detailed in Section~\ref{sec:convergence}. The evolutionary tracks of selected \texttt{MESA} models in the $\teff{}$ -- $\logg{}$ space are shown for comparison, as well as the final isochrone at $\qty{11.5}{Gyr}$. The tracks are labeled by their initial stellar masses in $\qty{}{\Msun}$.}
    \label{fig:teff_logg}
\end{figure}

PMS stars at zero age are initialized with uniform chemical abundances and central temperature of $T_c=\qty{5e5}{\K}$. This choice follows \citet{MIST} and \citet{roman_omega_cen}, and ensures that the stars of all considered initial masses are found within the boundary condition tables (Section~\ref{sec:bc}) at the completion of the main PMS relaxation routine in \texttt{MESA}. The additional PMS relaxation for $>\qty{0.3}{\Msun}$ stars until the formation of the radiative core (\texttt{pre\_ms\_relax\_to\_start\_radiative\_core}) was disabled in all of our evolutionary models to provide a consistent definition of zero age for all calculated evolutionary tracks. This feature was introduced in recent versions of the code to improve the robustness of massive star calculations\footnote{\href{https://github.com/MESAHub/mesa/issues/340}{https://github.com/MESAHub/mesa/issues/340}}, and does not significantly affect the final evolutionary states of the objects considered in this study.

The interior convective mixing length (in the formalism of \textit{mixing length theory}, MLT; \citealt{MLT}) in all models was set to the solar-calibrated value of $\alpha_\mathrm{MLT}=1.82$ scale heights \citep{solar_alpha,MIST}. The effect of other choices for this parameter is discussed in Section~\ref{sec:chemistry}. The input and output files for the evolutionary models produced in this study are made available \review{on our website}\footnote{\href{http://romanger.com/models.html}{http://romanger.com/models.html}} \review{and \dataset[Zenodo]{https://doi.org/10.5281/zenodo.10016008}}.

\begin{figure}[ht!]
    \centering
    \includegraphics[width=1\columnwidth]{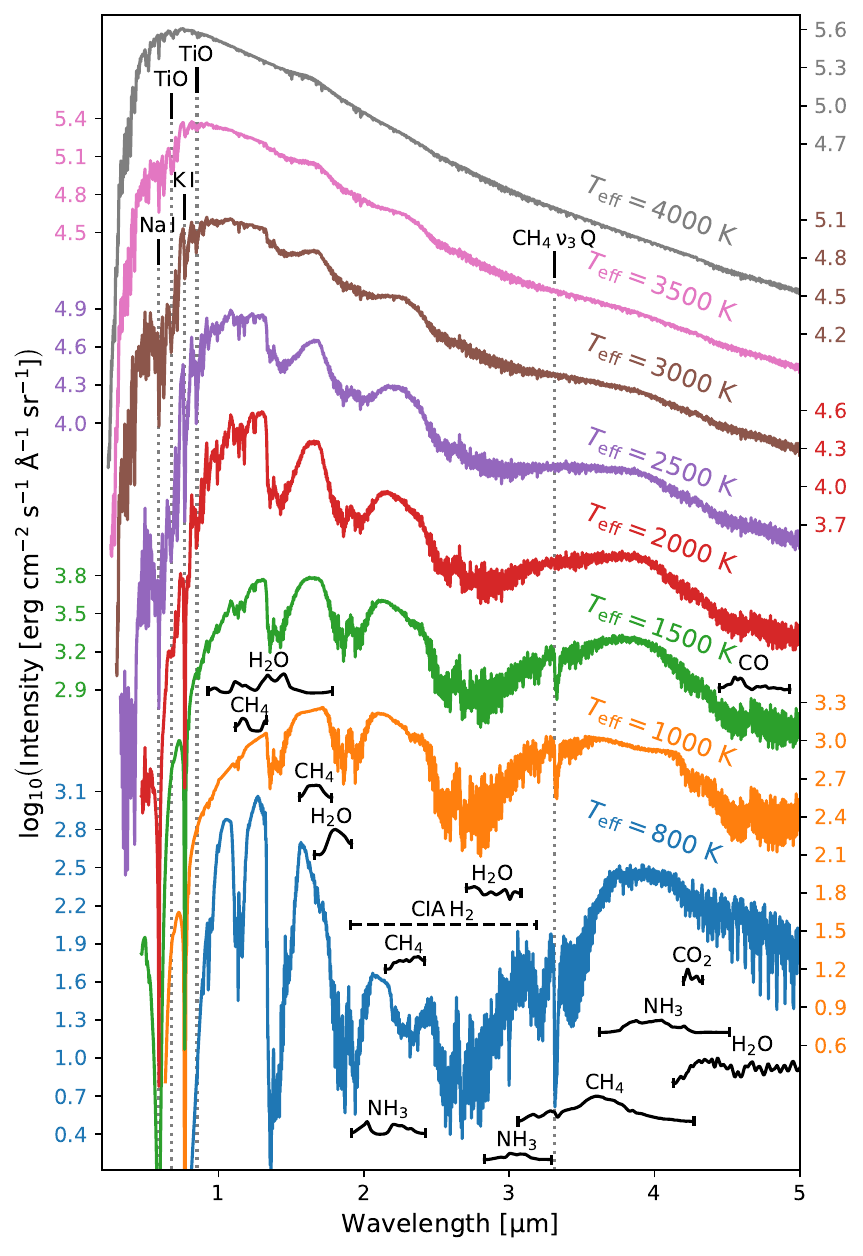}
    \caption{Synthetic spectra of selected model atmospheres at $\logg{}=5$, calculated in this study for the \textit{ridgeline} isochrone of \tuc{}. The important absorption bands of $\ce{CH_4}$, $\ce{CO}$, $\ce{CO_2}$, $\ce{H_2O}$ and $\ce{NH_3}$ are labeled with ragged black lines that represent the variation of absorption strength within the band for the $\teff{}=\qty{800}{\K}$ model. The central wavelengths of prominent $\ce{TiO}$ bands, \ce{Na} and \ce{K} atomic lines and the peak of the \ce{CH_4} $\nu_3$ (asymmetric stretching vibration band) Q-branch are shown with vertical dotted lines. The approximate wavelength range of strong collision-induced \ce{H_2} absorption (CIA) is shown with the horizontal dashed line. The spectra have been convolved with a $\qty{10}{\angstrom}$ Gaussian kernel for clarity.}
    \label{fig:spectra}
\end{figure}

\subsection{Boundary conditions}
\label{sec:bc}

At each evolutionary step, energy conservation requires the pressure-temperature profile to be consistent with the luminosity output of the star. Evaluating this condition near the surface of low-mass stars is challenging due to the complex relationship between the pressure-temperature profile and the outgoing energy flux, caused by low-temperature phenomena such as non-grey molecular opacity, cloud formation and convective instabilities. This issue is particularly prominent at $\teff{}\lesssim\qty{4500}{\K}$, as the structure of the stellar atmosphere begins to significantly deviate from the Eddington approximation \citep{integral_equation,grey_BC_1,grey_BC_2,sky_47_tuc}. More accurate atmospheric pressure-temperature profiles may be extracted from model atmospheres, precomputed for the entire range of surface parameters that may be encountered by the star during its evolution.

Following \citet{BC_origin,MIST,ATMO_boundary_conditions} and our earlier study \citep{roman_omega_cen}, we calculated a grid of model atmospheres (Section~\ref{sec:atmospheres}) for each theoretical isochrone, spanning $2\leq\logg{}\leq6$ for $\qty{2500}{\K}\leq\teff{}\leq\qty{7500}{\K}$, and $4\leq\logg{}\leq6$ for $\qty{500}{\K}\leq\teff{}\leq\qty{2400}{\K}$, with steps of $\qty{100}{\K}$ in $\teff{}$ and $0.5$ in $\logg{}$. Models with $\logg{}<4\ \&\ \teff{}<\qty{2500}{\K}$ were not required, since such low gravities are only encountered during the early evolution ($<\qty{5}{Myr}$) and the subgiant phase, both of which are characterized by higher effective temperatures ($\teff{}>\qty{2500}{\K}$). For some of the calculated isochrones, a small number of model atmospheres could not be computed due to convergence issues (Section~\ref{sec:convergence}). The corresponding empty grid points were then filled in using linear Delaunay triangulation \citep{qhull}. The temperatures and gravities of the atmosphere models calculated for the final \textit{ridgeline} isochrone are plotted in Figure~\ref{fig:teff_logg} alongside the $\teff{}$ -- $\logg{}$ tracks of selected evolutionary models.

The model atmospheres were converted to tables of gas pressure and temperature at a prescribed depth and provided to \texttt{MESA} as the outer boundary conditions at each evolutionary step and for each iteration in the solution of the stellar structure equations. In general, specifying the boundary conditions at larger Rosseland optical depths ($\tau$) is preferred, as it minimizes the discontinuity in opacity between the atmosphere and interior, and ensures adiabatic behavior at the boundary \citep{BC_origin,BC_origin_2,YaPSI}. However, the reduced atmospheric opacity in more massive stars shifts the boundary to larger physical depths, where the non-ideal gas behavior may be significant \citep{tau_100_vs_photosphere}, and is not accounted for in our high-$\teff{}$ model atmospheres. We therefore employed two distinct atmosphere-interior coupling regimes with boundary conditions at $\tau=100$ for $M<\qty{0.5}{\Msun}$, and at $T(\tau)=\teff{}$ (photosphere) for $M>\qty{0.6}{\Msun}$, where $T(\tau)$ is the temperature profile of the star and $M$ is the initial stellar mass. At intermediate stellar masses, $\qty{0.5}{\Msun}\leq M \leq \qty{0.6}{\Msun}$, a linear ramp between the two regimes was applied. The range of transition was chosen to approximately coincide with the maximum value of $dT(\tau=100)/dM$ for main sequence stars. The implications of this choice on the synthetic photometry are discussed in Section~\ref{sec:transitions}.

\review{When the range of pre-tabulated boundary conditions is exceeded, \texttt{MESA} implements a fail-safe and falls back on the Eddington approximation.} Near the edges of the table, a smooth blending between the on-table and off-table boundary conditions is carried out, which may result in numerical instabilities at low $\teff{}$ due to the extreme inaccuracy of the Eddington formula. Since our models, \review{by design}, do not leave the table range during regular evolution, we modified the source code of \texttt{MESA} to disable the off-table blending once the PMS phase has been completed. We also updated the code to interpolate the tables linearly rather than bicubically to address unphysical temperatures and pressures resulting from spline overshoots \citep{spline_overshoot} in the vicinity of poorly converged model atmospheres (Section~\ref{sec:convergence}). The calculated boundary condition tables and the patch file for the \texttt{MESA} codebase are made available \review{on our website}\footnote{\href{http://romanger.com/models.html}{http://romanger.com/models.html}} \review{and \dataset[Zenodo]{https://doi.org/10.5281/zenodo.10016008}}.

\subsection{Model atmospheres}
\label{sec:atmospheres}

Model atmospheres were calculated using the \texttt{PHOENIX} code \citep{phoenix_origin}, version \texttt{15.05.05D} \citep{BT-Settl,phoenix_15,roman_note} at $\teff{}\leq\qty{5000}{\K}$; and the \texttt{ATLAS} code \citep{ATLAS5}, version \texttt{9} \citep{ATLAS9_1,ATLAS9_2} at $\teff{}>\qty{5000}{\K}$. The use of \texttt{ATLAS} for ``warm'' atmospheres is advantageous due to its superior computational efficiency (see Section~\ref{sec:computational_cost}), attained by virtue of opacity sampling from pre-tabulated opacity distribution functions (ODFs; \citealt{ODFs_1,ODFs_2}), simplified chemical equilibrium (only gaseous species are considered, molecule-molecule interactions are ignored) and the ideal equation of state. The ODFs for the chemical composition of each isochrone were computed for $\qty{9}{\nano\meter}\lesssim\lambda\lesssim\qty{160}{\micro\meter}$ at $57$ temperatures between $\approx\qty{2e3}{\K}$ and $\approx\qty{2e5}{\K}$ using the satellite package \texttt{DFSYNHTE} \citep{DFSYNHTE,new_ODFs}. \texttt{ATLAS} atmospheres are stratified into $72$ plane-parallel layers, spanning $\sim\num{e-7}<\tau<\num{100}$ with logarithmic spacing. The synthetic spectra for each model atmosphere were calculated using the \texttt{SYNTHE} code \citep{SYNTHE} within $\qty{0.1}{\micro\meter}<\lambda<\qty{5}{\micro\meter}$, at the constant resolution\footnote{\texttt{SYNTHE} treats all line profiles as symmetric with respect to the nearest point on the wavelength grid. For this reason, high resolutions of order $\lambda/\Delta\lambda\sim10^5-10^6$ are typically recommended even when not required for intended science \citep{ATLAS9_2,cookbook}.} of $\lambda/\Delta\lambda\sim\num{6e5}$. All three codes are packaged in the open source \textit{Python} dispatcher \texttt{BasicATLAS}\footnote{\href{https://github.com/Roman-UCSD/BasicATLAS}{https://github.com/Roman-UCSD/BasicATLAS}} \citep{mikaela}, alongside a suite of internal consistency checks and a synthetic photometry calculator.

\texttt{PHOENIX} allows for a more accurate treatment of low-$\teff{}$ atmospheres through direct sampling of opacity at the wavelengths of interest, a more complete chemical reaction network (including condensation, see \citealt{PHOENIX_chemical_equilibrium} for a review), a comprehensive molecular line database (most importantly, \ce{H_2O} lines from \citealt{PL_BT} and \ce{CH_4} lines from \citealt{PL_LB_HITRAN} and \citealt{PL_STDS1,PL_STDS2}; other included lines are listed in Table~3 of \citealt{roman_omega_cen}), and a non-ideal equation of state \citep{PHOENIX_EOS}, among other features. The inclusion of condensate species (\textit{clouds}) in the chemical equilibrium and the opacity model is necessary to reproduce the observed reddening of photometric colors at $\teff\lesssim\qty{3000}{\K}$ compared to gas-only models \citep{PHOENIX_chemical_equilibrium,dust_formation}. Our setup of \texttt{PHOENIX} also includes the \textit{Allard \& Homeier} treatment of gravitational settling for condensate species \citep{phoenix_15,cloud_model_comparison} that gradually removes clouds from the spectrum-forming region of the atmosphere at $\teff\ll\qty{2000}{\K}$ \citep{rainout_1}. For computational efficiency, we only consider gravitational settling at $\teff<\qty{2500}{\K}$ and, otherwise, assume that the condensate species remain at chemical equilibrium.

Our \texttt{PHOENIX} models are stratified into $250$ layers when gravitational settling is used and $128$ layers otherwise. Unlike \texttt{ATLAS}, the atmospheric layers in \texttt{PHOENIX} are spherical and defined by the optical depth at a fixed wavelength ($\tau_\lambda$, where $\lambda=\qty{1.275}{\micro\meter}$) instead of the Rosseland mean\footnote{\citet{roman_omega_cen} have overlooked this distinction in definitions and, therefore, the boundary condition tables used in that study may be less accurate, especially at very low $\teff{}$, where the opacity at $\lambda=\qty{1.275}{\micro\meter}$ is no longer representative of the average across all wavelengths.}. The bottom of the atmosphere is set to $\tau_\lambda=\num{e3}$ for the models without gravitational settling, but reduced to $\tau_\lambda=316$ when settling is enabled to avoid the numerical issues associated with settling calculations at large depths. The top of the atmosphere is defined by the gas pressure, rather than optical depth, set to $\qty{e-3}{dyn\per\cm^2}$. The optical depth grid is approximately logarithmic. For all models considered in this study, the atmospheres at the final evolutionary state are negligibly thin compared to the stellar radius (the $\tau=100$ layer is always found within the outer $0.1\%$ of the photospheric radius, calculated at $T(\tau)=\teff{}$). Therefore, the effect of spherical geometry is expected to be subdominant, allowing the use of approximate atmospheric radii from literature \citep{radius_isochrones} instead of introducing additional dimensions to our model atmosphere grids. The wavelength sampling and spectral synthesis for all \texttt{PHOENIX} models are carried out within $\qty{1}{\angstrom}<\lambda<\qty{1}{\milli\meter}$ with a median resolution of $\lambda/\Delta\lambda\sim\num{18250}$ between $\qty{0.5}{\micro\meter}$ and $\qty{3}{\micro\meter}$. A few examples of the synthetic spectra for the chemical abundances of the \textit{ridgeline} isochrone and $\logg{}=5.0$ are provided in Figure~\ref{fig:spectra}. The atmosphere models calculated in this study are made available online\footnote{\href{http://romanger.com/models.html}{http://romanger.com/models.html}}.

We assume that the chemical composition of the atmosphere is unaffected by stellar evolution and is identical to the initial composition of the PMS star, with the exception of \ce{[Li/Fe]} that is reduced by $\qty{3}{dex}$ in all \texttt{PHOENIX} models compared to the PMS abundance. This assumption is justified by the fact that low-mass stars ($\lesssim\qty{0.6}{\Msun}$) undergo minimal nuclear processing (with the exception of lithium burning), while their higher-mass counterparts establish radiative zones at young ages, thereby preventing the nuclear burning products from reaching the atmosphere. The behavior of \ce{[Li/Fe]} in the interior and the atmosphere of the star as a function of mass is explored in detail in \citet{roman_omega_cen}.

\subsection{Model convergence}
\label{sec:convergence}

In both model atmospheres and the time steps of evolutionary models, the equations of input physics are solved using iterative methods. At the high effective temperatures of \texttt{ATLAS} atmospheres, the parameter space maintains approximate local linearity \citep{MARCS}, allowing for fast and reliable convergence onto a self-consistent solution. We carry out \texttt{ATLAS} iterations in batches of $15$ until the maximum flux error and the maximum flux derivative error across all layers drop below $1\%$ and $10\%$, respectively  (following \citealt{cookbook,convergence_standard}; see the appendix in \citealt{mikaela} for details), or until no further progress can be made. These convergence standards have been met by nearly all \texttt{ATLAS} atmospheres computed in this study, with the exception of a few $\teff{}>\qty{6500}{\K}$ models that generally fall outside the region of parameter space, explored by the calculated evolutionary tracks (Figure~\ref{fig:teff_logg}).

The evolution of convergence parameters across iterations in low-$\teff{}$ \texttt{PHOENIX} atmospheres is more involved. In models with subdominant cloud settling ($\teff{}\gtrsim\qty{2000}{\K}$), the flux errors generally decrease with every iteration until the model arrives at a stable solution with the typical maximum and average flux errors across all radiative layers of $\sim3\%$ and $0.5\%$, correspondingly. In this work, we refer to this convergence pattern as \textit{stable}, and the models that exhibit this pattern are shown in \textit{green} in Figure~\ref{fig:teff_logg} for the \textit{ridgeline} isochrone. A small number of models with stable convergence may occasionally encounter iterations with ill-conditioned temperature corrections that break the trend in convergence and increase the flux errors by over an order of magnitude, effectively restarting the computation. In those cases, the iteration with the minimum average error, from which we derive the final synthetic spectrum, is still expected to produce a reliable result; however, the gain in accuracy from an increased number of iterations may be limited.

At lower $\teff{}$, the effects of condensate settling become more prominent, resulting in a far more complex relationship between the model parameters that restricts the effectiveness of the temperature correction scheme. In this regime, the flux errors typically oscillate between high and low values across iterations. While the flux errors of the best iteration are usually comparable to those of the atmospheres with stable convergence, the final solutions to the structure equations may lack uniqueness (i.e. for some $\teff{}$ and $\logg{}$ there may be multiple atmosphere structures with similar flux errors but vastly different emergent spectra). Furthermore, the pronounced sensitivity of the model to temperature corrections casts doubt on the usefulness of the radiative flux errors as a diagnostic of self-consistency. We refer to the convergence pattern in this temperature regime as \textit{irregular}. Some of the \texttt{PHOENIX} atmospheres with irregular convergence had to be excluded from the grid due to the rapid growth of the oscillation amplitude in the temperature structure between iterations. For the rest of models with this convergence pattern, the solution with the lowest average radiative flux error was added to the grid at the points identified in Figure~\ref{fig:teff_logg} with \textit{black} markers for the \textit{ridgeline} isochrone.

A small number of models displayed a convergence pattern, intermediate between that of stable convergence and irregular convergence, which we refer to as \textit{limited} convergence. In those cases, the model may exhibit stable convergence for the first few iterations, but then transition into the irregular convergence pattern before a solution with satisfactory flux errors can be reached. In some cases, the transition between the two convergence patterns may occur multiple times over the course of $\sim 100$ iterations. The models with limited convergence are shown in Figure~\ref{fig:teff_logg} with \textit{red} markers for the \textit{ridgeline} isochrone. At very low temperatures ($\teff{}\lesssim\qty{1200}{\K}$) and high gravities ($\logg{}\gtrsim 5.5$), the condensate species are almost completely removed from the spectrum-forming region of the atmosphere, restoring the stable convergence pattern (see Figure~\ref{fig:teff_logg}).

For most \texttt{PHOENIX} models computed in this study, we used the nearest atmosphere structure from the \texttt{NEXTGEN} model grid \citep{nextgen,PHOENIX_simulator} as the initial guess in the solver, with the exception of models with particularly poor convergence that were recalculated, using the nearest well-converged atmospheres from our grid as the initial models instead. In principle, the accuracy of our low-temperature atmospheres may be improved by calculating the models in small batches with progressively decreasing $\teff{}$ and $\logg{}$, and using the atmospheres from the preceding batch as initial guesses. In fact, we adopted this approach for the \texttt{ATLAS} models. However, the high computational demand of \texttt{PHOENIX} (Section~\ref{sec:computational_cost}) makes the un-parallelized computation extremely time-consuming. Furthermore, it is unclear how the potential gain in numerical precision would compare to the systematic errors in the input physics and grid interpolation errors. We therefore chose to focus on identifying the outlier models based on their synthetic photometry and excluding them from the isochrone calculation instead, as described in Section~\ref{sec:synphot}.

Nearly all evolutionary \texttt{MESA} models in our grids have converged at every time step within the ``\textit{gold tolerances}'' \citep{MESA_5}, with the exception of a few lowest-mass models ($\lesssim\qty{0.05}{\Msun}$), where the tolerances were relaxed to their standard values to allow evolution over the discontinuities in $dT(\tau=100)/d\logg{}$, caused by the sparse gravity sampling in the boundary condition tables.

\subsection{Computational demand of model atmospheres}
\label{sec:computational_cost}

All model atmospheres used in this work were calculated on the \textit{Bridges-2} supercomputer \citep{Bridges-2}, operated by the Pittsburgh Supercomputing Center. Our \texttt{PHOENIX} models were calculated to $\sim 100$ iterations, requiring between $\sim0.15$ processor hours per iteration at the highest temperatures to $\sim 1.5$ processor hours at the lowest. The high memory demand of \texttt{PHOENIX} ($\gtrsim\qty{4}{GB}$ per model) necessitated requesting at least $2$ (and occasionally $3$) service units for each processor hour.

The computational demand of \texttt{ATLAS} models is dominated by the spectral synthesis, carried out by the \texttt{SYNTHE} package. The calculation of the emergent spectra for most \texttt{ATLAS} models required between $1.5$ and $3.5$ processor hours per spectrum (and the same number of service units). For comparison, the structure calculations only took $\approx0.005$ processor hours per model for all iterations.

The computational demand of a complete atmosphere grid for each chemical composition is approximately $\num{5e4}$ service units, with nearly $98\%$ taken by \texttt{PHOENIX} calculations. This estimate does not include recalculation of failed models, ODF calculations, calculations of partial pressure tables for each chemical composition used in \texttt{PHOENIX}, and evolutionary calculations with \texttt{MESA}. This estimate only applies to the final isochrones (\textit{ridgeline}, \textit{blue tail} and \textit{red tail}) presented in Section~\ref{sec:chemistry}: to calculate the intermediate isochrones used during the fitting process, we took advantage of various optimizations described in Section~\ref{sec:chemistry}.

\subsection{Isochrone stitching}
\label{sec:transitions}

The process of calculating theoretical isochrones described earlier in this section required a choice of three ``stitching points'', where distinct modelling setups were blended together: the effective temperature where \texttt{ATLAS} models are replaced with \texttt{PHOENIX} models ($\qty{5000}{\K}$), the effective temperature where the gravitational settling in \texttt{PHOENIX} models was disabled ($\qty{2500}{\K}$) and the initial stellar mass, where the $\tau=100$ boundary condition tables were replaced with $T(\tau)=\teff{}$ (smooth ramp between $\qty{0.5}{\Msun}$ and $\qty{0.6}{\Msun}$). Here, we briefly review the implications of those choices for synthetic photometry.

Our choice of the \texttt{ATLAS}/\texttt{PHOENIX} transition temperature ($\qty{5000}{\K}$) is higher than that used in our previous work ($\qty{4000}{\K}$, \citealt{roman_omega_cen}). For comparison, the published grid of \texttt{ATLAS} models with the updated molecular opacities \citep{new_ODFs} reaches $\qty{3500}{\K}$, although the low-temperature extension has known issues \citep{ATLAS_eval}. We calculated an additional test set of \texttt{ATLAS} models for the parameters of the \textit{ridgeline} isochrone at $\qty{4000}{\K}\leq\teff{}\leq\qty{5000}{\K}$, as well as an additional grid of evolutionary models with the updated boundary conditions. We found the synthetic photometry at $\qty{11.5}{Gyr}$ of the regular \textit{ridgeline} isochrone and its altered version to be indistinguishable at $\teff=\qty{5000}{\K}$, but rapidly diverging at lower temperatures to attain the difference of $\approx \qty{0.025}{mag}$ in \texttt{F606W} -- \texttt{F814W} and \texttt{F150W2} -- \texttt{F322W2} at $\teff{}=\qty{4000}{\K}$. The difference originates almost entirely from the synthetic spectra with only a minor contribution from the boundary conditions.

Our choice of the minimum $\teff{}$ without gravitational settling ($\qty{2500}{\K}$) is, on the other hand, lower than that employed in our previous study of \omegacen{} ($\qty{3000}{\K}$, \citealt{roman_omega_cen}), which allowed us to significantly reduce the computational demand. We calculated a test grid of \texttt{PHOENIX} atmospheres for the \textit{ridgeline} isochrone with $\qty{2500}{\K}\leq\teff{}\leq\qty{3000}{\K}$ and enabled settling of condensates, as well as a test set of corresponding evolutionary models. We found that at $\teff{}=\qty{2500}{\K}$ and at the age of $\qty{11.5}{Gyr}$, the effect of gravitational settling on the same colors as before is $\approx\qty{0.01}{mag}$. In the optical regime, both the boundary conditions and the synthetic spectra make comparable contributions to the observed difference, while the latter dominate in the infrared. At $\teff{}>\qty{2500}{\K}$, the effect of gravitational settling decreases rapidly in the infrared, but remains approximately constant in the optical up to the warm end of the test grid at $\qty{3000}{\K}$, primarily due to the effect of the updated boundary conditions on stellar evolution.

Lastly, we examine the effect of the transition between the atmosphere-interior coupling schemes by disabling the linear ramp at $\qty{0.5}{\Msun}\leq M \leq \qty{0.6}{\Msun}$. The effect on synthetic photometry at $\qty{11.5}{Gyr}$ is most prominent in \texttt{F606W} -- \texttt{F814W} when the range of $T(\tau)=\teff{}$ boundary condition tables is extended down to $\qty{0.5}{\Msun}$, resulting in the discontinuity of $\lesssim\qty{0.01}{mag}$ between the two regimes. The discontinuity is less prominent when the $\tau=100$ tables are extended to $\qty{0.6}{\Msun}$ or when other photometric colors are considered.

In summary, our choice of ``stitching points'' in the isochrone is not expected to introduce errors larger than $\qty{0.01}{mag}$ in any of the photometric colors considered in this study; however, it appears that the optimal choice of the transition between the two atmosphere-interior coupling schemes falls at somewhat larger masses, and a small gain in accuracy may be attained by allowing gravitational settling of condensates at $\teff>\qty{2500}{\K}$.

\begin{figure}[ht!]
    \centering
    \includegraphics[width=1\columnwidth]{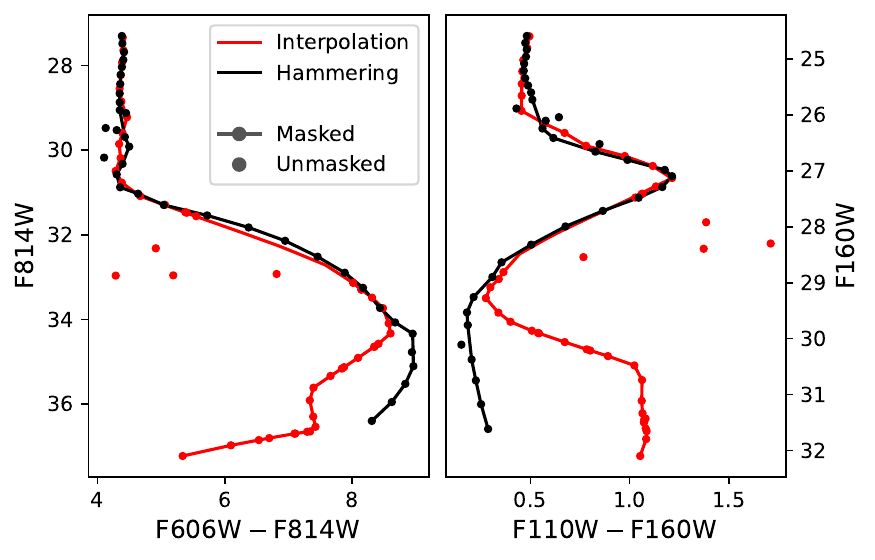}
    \caption{Comparison of synthetic photometry at \qty{11.5}{Gyr} and $\teff{}\leq\qty{2350}{\K}$ for the \textit{blue tail} isochrone, calculated using the two methods described in Section~\ref{sec:synphot}: atmosphere hammering (\textit{black}) and linear interpolation of the bolometric corrections (\textit{red}). The outlier models were identified by visual inspection in each isochrone and excluded from the ``masked'' result.}
    \label{fig:hammering}
\end{figure}

\subsection{Synthetic photometry and hammering}
\label{sec:synphot}
The preliminary synthetic photometry for the calculated isochrones was obtained by applying the reddening law to the synthetic spectra, evaluating the bolometric corrections for each model atmosphere in the bandpasses of interest, and interpolating the result to $\teff{}$ and $\logg{}$, predicted by the evolutionary model at the required initial mass and age. The resulting magnitudes were corrected by the distance modulus of $\qty{13.2418\pm0.0625}{mag}$ \citep{dm}. To calculate the bolometric corrections, we used the \texttt{synphot()} routine of the \texttt{BasicATLAS} package \citep{mikaela}. The process uses the reddening law from \citet{extinction}, parameterized only by the optical reddening, $E(B-V)$, and assuming the total-to-selective extinction ratio of $R_V=3.1$.

The strong dependency of low-$\teff{}$ spectra on surface parameters and the highly non-linear behavior of cool atmospheres pose a challenge to the interpolation of bolometric corrections. Linear interpolation of a sparsely sampled grid leads to large unphysical discontinuities in the derivatives, and is known to introduce noticeable errors even at relatively high temperatures \citep{high_order_interpolation,MARCS}. On the other hand, higher-order interpolation methods are less robust against outliers \citep{spline_overshoot}, such as those introduced by the models with poor convergence. Specialized interpolation methods for atmosphere grids with non-convergent models (e.g. \citealt{broken_model_grids}) generally require a means to reliably identify the outliers, which is not straightforward in the case of irregular convergence, as discussed in Section~\ref{sec:convergence}.

To address these issues, we decided to avoid interpolating the grid at low temperatures altogether and, instead, to calculate additional \texttt{PHOENIX} atmospheres at $\qty{600}{\K}\leq\teff{}<\qty{2400}{\K}$ in $\qty{50}{\K}$ steps, with $\logg{}$ set to the values predicted by the isochrone for each effective temperature at the target age. We refer to this process as atmosphere \textit{hammering}. In addition to removing the need to interpolate bolometric corrections at low temperatures, hammering serves two other purposes. First, it reduces the effective number of dimensions in the atmosphere grid from $2$ to $1$ (since the hammering models obey a known $\teff{}$ -- $\logg{}$ relationship), allowing the outlier models to be easily identified, e.g., by their placement in the color-magnitude space. Second, it allows us to derive the synthetic photometry at $\teff{}<\qty{2500}{\K}$ from higher-gravity model atmospheres that have better convergence (e.g., from Figure~\ref{fig:teff_logg}, to calculate the bolometric corrections at the $\teff{}=\qty{2000}{\K}$ point on the $\qty{11.5}{Gyr}$ isochrone, one would require both $\logg{}=5.0$ and $\logg{}=5.5$ model atmospheres for the interpolation method, but only one $\logg{}\approx5.5$ model for the hammering method). The major drawbacks of the hammering method are the added computational cost and the resulting commitment to a particular target age, since the hammering models fix the $\teff{}$ -- $\logg{}$ relationship.

The linear interpolation and hammering methods of calculating synthetic photometry at $\teff{}\leq\qty{2350}{\K}$ are compared in Figure~\ref{fig:hammering}. The figure demonstrates that at $\teff{}\lesssim\qty{1000}{\K}$ (the brown dwarf regime), linear interpolation may be off by as much as $\qty{2}{mag}$ in the optical regime and $\gtrsim\qty{0.5}{mag}$ in the near infrared. It is also apparent that the two methods converge at $\teff{}\gtrsim\qty{2000}{\K}$, suggesting that linear interpolation of bolometric corrections is valid across the main sequence. In this study, we use linear interpolation for all models with $\teff{}>\qty{2350}{\K}$.

We note that while the synthetic photometry inferred from atmosphere hammering does not rely on interpolated bolometric corrections, it is still based on the surface parameters predicted by the evolutionary track, which, in turn, was calculated using interpolated boundary conditions. The effect of boundary conditions on synthetic photometry is subdominant compared to that of bolometric corrections; however, some interpolation errors may remain at low $\teff{}$. To avoid interpolation of model atmospheres entirely, one must calculate new atmospheres for every iteration of the evolutionary code at every time step, which would increase the total computational demand by over an order of magnitude.

\section{Isochrone fitting} \label{sec:chemistry}
\begin{figure}[ht!]
    \centering
    \includegraphics[width=1\columnwidth]{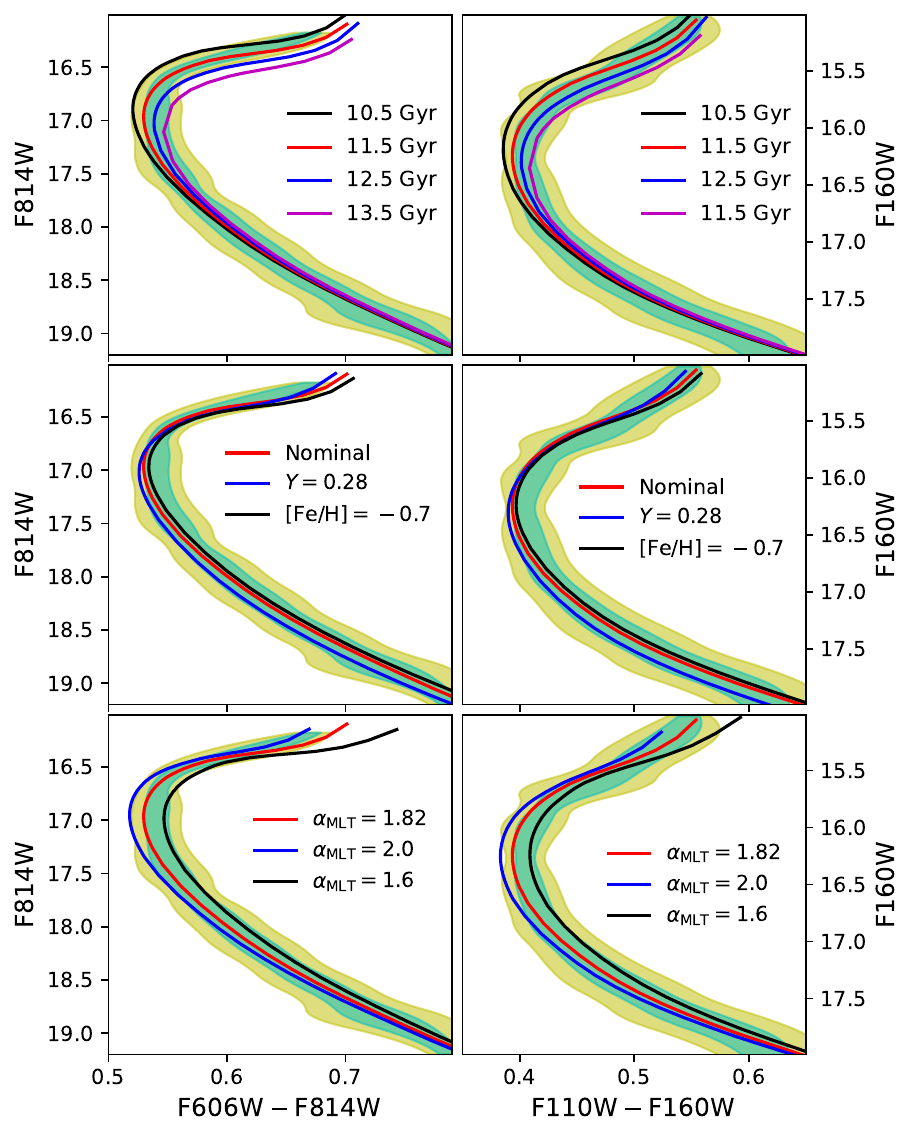}
    \caption{\textit{Top:} \textit{ridgeline} isochrones of \tuc{} at different ages, overplotted on the distribution of observed colors in optical (\textit{left}) and near infrared (\textit{right}). \textit{Middle}: comparison of the \textit{ridgeline} isochrone at $\qty{11.5}{Gyr}$ with the nominal \ce{[Fe/H]} and $Y$ (\textit{red}) to the helium-rich (\textit{blue}) and metal-rich (\textit{black}) isochrones at the same age. \textit{Bottom:} effect of the mixing length parameter on the \textit{ridgeline} isochrone at the age of $\qty{11.5}{Gyr}$. The cyan and yellow regions of the color-magnitude diagram subtend the $1\sigma$ and $2\sigma$ contours in the data distribution, respectively. $E(B-V)=\qty{0.04}{mag}$ for all shown isochrones.}
    \label{fig:turnoff}
\end{figure}

\subsection{Main sequence turn-off and subgiant branch}

The high effective temperatures of the main sequence turn-off and the subgiant branch in \tuc{} ($\teff{}\gtrsim\qty{5000}{\K}$) suppress molecular formation in the atmosphere and minimize the impact of chemical abundances on the observed photometry. The notable exceptions are the overall metallicity, \ce{[Fe/H]}, and the helium mass fraction, $Y$, as they have a significant effect on the mean molecular weight and opacity of the interior that, in turn, guide the evolutionary track of the star \citep[Ch. 5.5]{salaris_textbook}. These high-mass stars are also sensitive to the adopted energy transport parameters (in particular, the mixing length, $\alpha_\mathrm{MLT}$) due to the emergence of an outer convective zone \citep{not_all_stars_are_the_sun}. To reduce the degree of degeneracy in the determination of chemical abundances, it is advantageous to constrain as many parameters as possible from the main sequence turn-off and the subgiant branch before analyzing the lower main sequence. We used this part of the CMD to assess our choice of $\alpha_\mathrm{MLT}=1.82$, fix the cluster age and the optical reddening, $E(B-V)$, as well as to evaluate the accuracy of the helium mass fraction and metallicity in the nominal chemical composition (Section~\ref{sec:nominal_chem}, $Y=0.25$, $\ce{[Fe/H]}=-0.75$).

The upper panels of Figure~\ref{fig:turnoff} show the turn-off point and the subgiant branch of the \textit{ridgeline} isochrone, evaluated at $4$ distinct ages and overplotted on the color-magnitude distribution of the archival photometry (Section~\ref{sec:data}) in the optical (\textit{left}) and near infrared (\textit{right}). The isochrones were calculated using the nominal metallicity and helium mass fraction, as well as $E(B-V)=\qty{0.04}{mag}$, in agreement with \citet{reddening}. In the figure, the $\qty{11.5}{Gyr}$ isochrone is the only one that matches both the color of the turn-off point and the luminosity of the subgiant branch within one standard deviation. Furthermore, the presented fit firmly fixes the reddening at $\qty{0.04}{mag}$, as lower values would result in the isochrone being ``too blue'' at the turn-off point, while higher values would make the isochrone ``too faint'' at the tip of the subgiant branch. This age estimate broadly agrees with the literature (e.g. \citealt{47_Tuc_age}; also see \citealt{47_Tuc_age_review} for a compilation of recent age measurements).

To explore the effect of the helium mass fraction and the overall metallicity, we calculated two additional isochrones with $Y=0.28$ and $\ce{[Fe/H]}=-0.7$, with all other parameters matching those of the \textit{ridgeline} isochrone. The results are plotted in the middle panels of Figure~\ref{fig:turnoff}. While the enhanced metallicity offers a marginal improvement of the fit at the turn-off point, the corresponding decrease in luminosity of the subgiant branch reduces the overall goodness of fit. On the other hand, the higher helium mass fraction improves the fit at the subgiant branch, at the expense of mismatching the color of the turn-off point by nearly two standard deviations in near infrared. Based on these results, we chose to use the nominal metallicity and helium mass fraction for all isochrone calculations henceforth.

The lower panels of the figure demonstrate how our choice of the mixing length compares to two alternative values: $\alpha_\mathrm{MLT}=1.6$ and $\alpha_\mathrm{MLT}=2.0$. The isochrones are shown at $\qty{11.5}{Gyr}$. Around the turn-off point, the effect is practically indistinguishable from that of the cluster age, emphasizing the limitations of using the color of the turn-off point as an age diagnostic for globular clusters. Our earlier claim that the $\qty{11.5}{Gyr}$ isochrone is the only one that fits the observed scatter within one standard deviation is invalid if the mixing length is allowed to vary as a free parameter. We found that a comparable goodness-of-fit is obtained at $\qty{12.5}{Gyr}$ for $\alpha_\mathrm{MLT}=2.0$, or $\qty{10.5}{Gyr}$ for $\alpha_\mathrm{MLT}=1.6$. Figure~\ref{fig:turnoff} appears to indicate that the effect of the mixing length is most significant at the tip of the subgiant branch. If the trend continues into the red giant branch, it is possible that a better estimate of the cluster age may be obtained from higher-mass stars. Since the degeneracy cannot be broken within the mass range considered in this study, we chose to adopt the solar-calibrated $\alpha_\mathrm{MLT}=1.82$ for all isochrones and the corresponding best-fit age of $\qty{11.5}{Gyr}$.

\subsection{Lower main sequence}

Below the main sequence inflection point ($\sim\qty{0.5}{\Msun}$, \citealt{MS_inflection_1}), the abundances of individual chemical elements may noticeably impact the shape of the CMD. In this study, we consider the variations in $11$ element abundances: \ce{[C/Fe]}, \ce{[N/Fe]}, \ce{[O/Fe]}, \ce{[Na/Fe]}, \ce{[Mg/Fe]}, \ce{[Al/Fe]}, \ce{[Si/Fe]}, \ce{[S/Fe]}, \ce{[Ca/Fe]}, \ce{[Ti/Fe]}, and \ce{[V/Fe]}.

The effect of atomic abundances is twofold. First, variations in abundances displace the chemical equilibrium of the atmosphere and change the corresponding opacity distribution, resulting in altered photometric colors. Second, the new atmospheric structure affects the boundary conditions of the atmosphere-interior coupling (Section~\ref{sec:bc}), thereby offsetting the end-point of the evolutionary track to different $\teff{}$ and $\logg{}$. The importance of the latter effect approximately correlates with $\delta\kappa=\left|d\kappa/d\ce{[X/Fe]}\right|$ at $T(\tau)=T_\mathrm{eff}$, where \ce{[X/Fe]} is the abundance of the element of interest and $\kappa$ is the Rosseland mean opacity. For the range $\qty{0.1}{\Msun}<M<\qty{0.5}{\Msun}$, this diagnostic is by far the largest for \ce{[O/Fe]} and \ce{[Ti/Fe]} ($\sim\qty{e-7}{cm^{-1}dex^{-1}}$ at $\qty{0.2}{\Msun}$) due to the prominent \ce{TiO} and \ce{H_2O} absorption bands (see Figure~\ref{fig:spectra}). For comparison, the next two most important elements, \ce{[C/Fe]} and \ce{[Al/Fe]}, have $\delta\kappa\sim\qty{2e-8}{cm^{-1}dex^{-1}}$ and $\sim\qty{e-8}{cm^{-1}dex^{-1}}$, respectively, at the same initial stellar mass.

The relative importance of the two effects depends on the chosen photometric bands. For instance, the optical main sequence spectra for fixed $\teff{}$ and $\logg{}$ are only weakly sensitive to the oxygen abundance, since \ce{H_2O} bands are mostly confined to infrared wavelengths, while the rate of \ce{TiO} production in the atmosphere is primarily determined by \ce{[Ti/Fe]}. As such, the atmosphere-interior coupling alone is responsible for over $50\%$ of the correlation between the optical colors (\texttt{F606W} -- \texttt{F814W}) and \ce{[O/Fe]} for main sequence stars. This example emphasizes that accounting for the enhancements of individual elements in the atmosphere-interior coupling scheme is essential for the accurate photometric determination of the chemical composition. Since \tuc{} is known to have a significant spread in \ce{[O/Fe]} based on spectroscopic measurements, the corresponding distribution of photometric colors cannot be captured with atmosphere-interior boundary conditions based on solar or even solar-scaled abundances. Furthermore, since both \ce{[Ti/Fe]} and \ce{[O/Fe]} have comparable $\delta\kappa$, and since both are considered to be $\alpha$-elements, even solar-scaled boundary conditions with adjustable $\alpha$-enhancement would not be adequate.

\subsection{Abundance variation grids}
\label{sec:abundance_variation_grids}

We determine the final \textit{red tail}, \textit{blue tail} and \textit{ridgeline} isochrones for \tuc{} by iteratively correcting the nominal element abundances, derived in Section~\ref{sec:nominal_chem}. Each iteration begins by calculating the theoretical isochrone with fully self-consistent chemistry as detailed in Section~\ref{sec:isochrones}. However, since the available datasets only extend to the end of the main sequence (Section~\ref{sec:data}), we terminate all intermediate isochrones at $\teff{}=\qty{3000}{\K}$ to avoid unnecessary calculations of model atmospheres.

\begin{figure}[ht!]
    \centering
    \includegraphics[width=1\columnwidth]{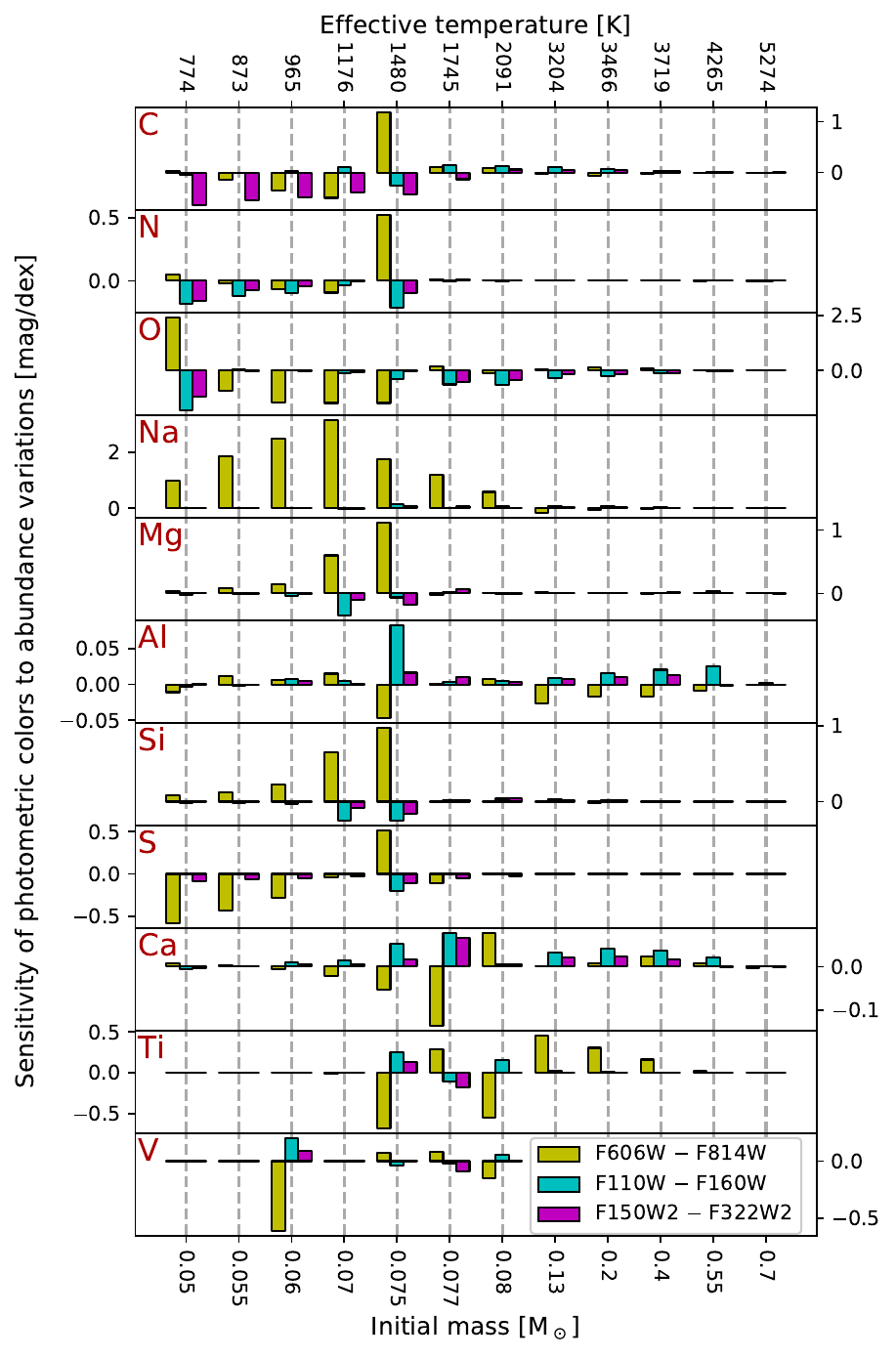}
    \caption{Photometric sensitivity plot for the initial iteration with nominal chemistry at $M>\qty{0.1}{\Msun}$ and for the final \textit{ridgeline} isochrone at $M<\qty{0.1}{\Msun}$. The plot shows the effect of changes in the abundances of individual elements on the photometric colors as a function of stellar mass (or, equivalently, $\teff{}$) assuming that the atmosphere-interior coupling remains constant, i.e. for fixed $\teff{}$ and $\logg{}$. Note that the horizontal axis is categorical, i.e. the labels are evenly spaced regardless of their values. The data shown in this figure are available as a machine-readable table in the digital version of this publication.}
    \label{fig:variations}
\end{figure}

To determine the corrections in the element abundances for the next iteration, we first compute an \textit{abundance variation grid} of model atmospheres for the current isochrone. The grid spans $5$ initial stellar masses between $\qty{0.13}{\Msun}$ and $\qty{0.7}{\Msun}$ that sample the main sequence of the cluster with approximately even intervals in the CMD. At this stage, we assume that the effect of chemistry on the atmosphere-interior coupling is negligible and compute new model atmospheres (Section~\ref{sec:atmospheres}) for the $\teff{}$ and $\logg{}$ of the chosen masses, based on the parameter relationships of the current isochrone. For each stellar mass, we calculate $22$ new model atmospheres with each of the $11$ elements considered in this study perturbed by $\qty{+0.5}{dex}$ and $\qty{-0.5}{dex}$, one element at a time. A \textit{photometric sensitivity plot} (Figure~\ref{fig:variations}) is then produced that shows the effect of each element on the relevant photometric colors. Since the abundance variation grids are based on the assumption of constant atmosphere-interior coupling, we also calculate $\delta\kappa$ for each element to estimate our confidence in the derived correlation between colors and abundances.

\begin{figure}[ht!]
    \centering
    \includegraphics[width=1\columnwidth]{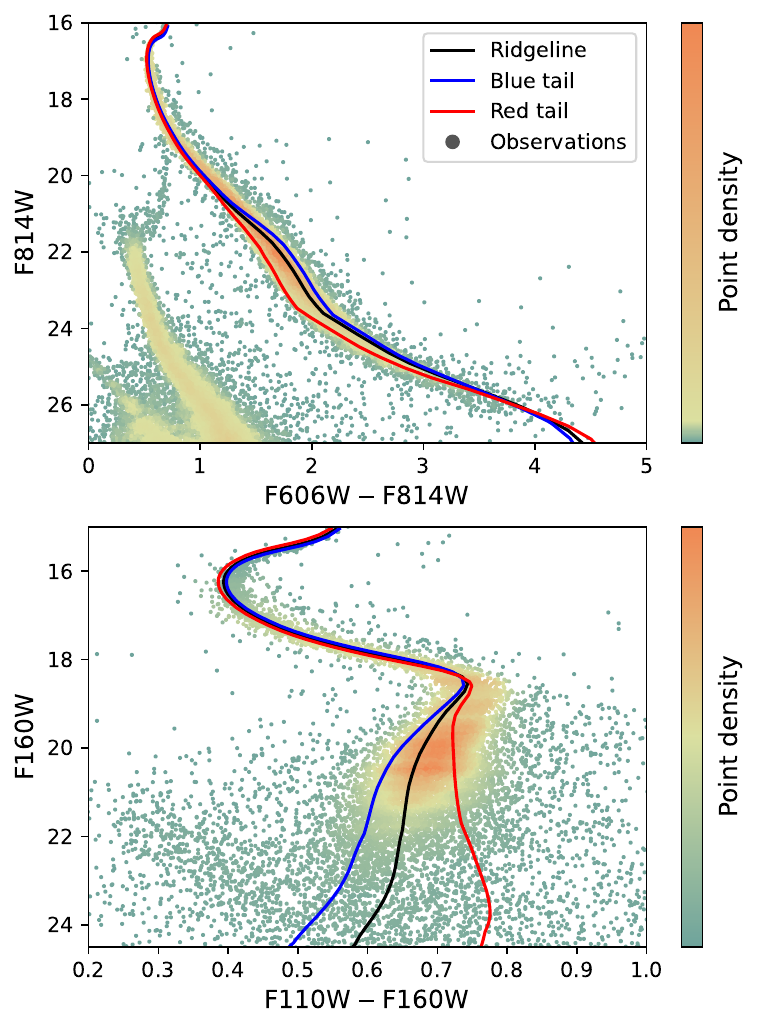}
    \caption{The final \textit{best fit}, \textit{blue tail} and \textit{red tail} isochrones, overplotted on the observed photometric spread in the optical (\textit{top}) and the near infrared (\textit{bottom}) bands. The parameters of the isochrones are given in Table~\ref{table:best_fit}. The isochrones shown in this figure are available as machine-readable tables in the digital version of this publication.}
    \label{fig:best_fit}
\end{figure}

The abundance corrections are then determined manually based on the current discrepancies between the isochrones and the data, the photometric sensitivity plot and estimated adjustments to the photometric sensitivity based on the $\delta\kappa$ values and our experience with earlier iterations. At each iteration, we follow the reductionist approach of adopting the smallest possible number of corrections in the chemical composition to reproduce the data. In principle, the derivation of abundance corrections may be automated by introducing a quantitative goodness-of-fit criterion, similar to the one derived in our previous work for fitting a single isochrone to the CMD \citep{roman_omega_cen}. However, since there is no straightforward way to extend that criterion to simultaneous fitting of multiple isochrones, and since the automated routine must be ``trained'' to take advantage of the abundance variation grids based on experience, we chose to use human input at every iteration instead.

A satisfactory fit was obtained after $10$ iterations. The final isochrones were extended to $\teff{}\gtrsim\qty{700}{\K}$ to reach the substellar regime, as required for our further analysis. For the final \textit{ridgeline} isochrone, an additional abundance variation grid was computed for $7$ initial masses below the end of the main sequence at $\qty{0.05}{\Msun}\leq M\leq\qty{0.08}{\Msun}$. The derived photometric sensitivity is shown in Figure~\ref{fig:variations}. We determined that corrections in \ce{[Ti/Fe]} and \ce{[O/Fe]} are sufficient to reproduce the photometric spread in both optical and infrared colors, while variations in other abundances compared to their spectroscopic means do not offer a noticeable improvement to the fit. While the value of \ce{[Ti/Fe]} needed to be offset from its nominal value to fit the data, we found that the observations are most consistent with a constant value of \ce{[Ti/Fe]} across all three final isochrones.

The final isochrones are plotted in Figure~\ref{fig:best_fit}, while their best-fit properties are summarized in Table~\ref{table:best_fit}. In near infrared bands, the photometric spread noticeably overflows the \textit{red tail} isochrone around the main sequence \review{knee} (\texttt{F160W} $\sim\qty{19}{mag}$). A similar red excess in the optical CMD is also seen at the corresponding evolutionary phase (\texttt{F814W} $\sim\qty{20.5}{mag}$), despite the inverted \ce{[O/Fe]}-color relationship. This prominent discrepancy between the data and the model isochrones cannot be accounted for by varying any of the considered elements.

\begin{deluxetable}{lccc}
\tablecaption{Best-fit parameters of the final isochrones \label{table:best_fit}}
\tablewidth{0pt}
\tablehead{
\colhead{Parameter} & \colhead{Blue tail} & \colhead{Ridgeline} & \colhead{Red tail}
}
\startdata
$E(B-V)$ [$\qty{}{mag}$]   &    $0.04$ &    $0.04$  &    $0.04$  \\
$\alpha_\mathrm{MLT}$   &    $1.82$ &    $1.82$  &    $1.82$  \\
Age [$\qty{}{Gyr}$]   &    $11.5$ &    $11.5$  &    $11.5$  \\
\ce{[Fe/H]} [$\qty{}{dex}$]   &    $-0.75$ &    $-0.75$  &    $-0.75$  \\
$Y$   &    $0.25$ &    $0.25$  &    $0.25$  \\
\ce{[Ti/Fe]} [$\qty{}{dex}$]   &    $0.64$ &    $0.64$  &    $0.64$  \\
\ce{[O/Fe]} [$\qty{}{dex}$]   &    $0.48$ &    $0.33$  &    $0.03$  \\
\hline
\enddata
\tablecomments{The abundances of individual elements unlisted in the table are set to their nominal values, derived in Section~\ref{sec:nominal_chem} and summarized in Appendix~\ref{sec:nominal_chem_table}.}
\end{deluxetable}

\section{Analysis}\label{sec:analysis}

    \subsection{Oxygen abundance}\label{sec:abundances}
    Figure~\ref{fig:best_fit} demonstrates that the near infrared photometry of the lower main sequence is particularly sensitive to the oxygen abundance of \tuc{}. Here, we calculate the lower main sequence distribution of \ce{[O/Fe]} in the cluster. Each of the three isochrones shown in the figure defines the relationship between the initial mass of the star and its position in the CMD space for a given oxygen abundance ($\ce{[O/Fe]}=\qty{0.48}{dex}$ for the \textit{blue tail} isochrone, $\ce{[O/Fe]}=\qty{0.33}{dex}$ for \textit{ridgeline} and $\ce{[O/Fe]}=\qty{0.03}{dex}$ for \textit{red tail}; see Table~\ref{table:best_fit}). Equivalent relationships for other values of \ce{[O/Fe]} may be approximated by interpolating or extrapolating the calculated isochrones. It is therefore possible to define a curvilinear transformation from the CMD space in Figure~\ref{fig:best_fit} to the $M-\ce{[O/Fe]}$ space.

\begin{figure}[ht!]
    \centering
    \includegraphics[width=1\columnwidth]{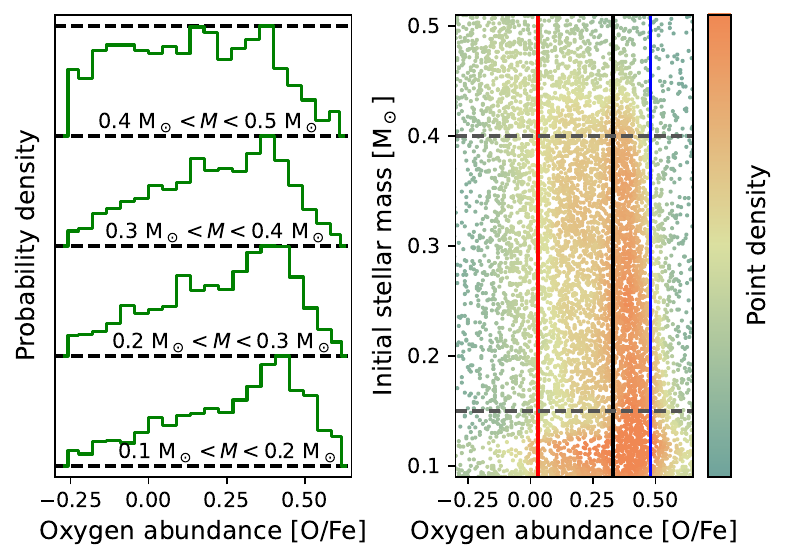}
    \includegraphics[width=1\columnwidth]{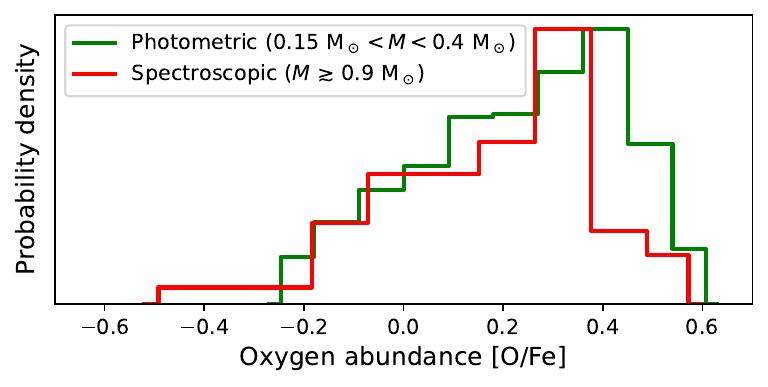}
    \caption{\textit{Left:} distribution histograms of \ce{[O/Fe]} in \tuc{} in different initial mass bins. The flattening of the highest-mass distribution is likely an artifact of the poor isochrone fit in the vicinity of the main sequence \review{knee} (see Section~\ref{sec:abundance_variation_grids}). The lowest-mass distribution is contaminated by SMC members. The horizontal dashed lines indicate the range of masses where the analysis of the oxygen distribution was carried out. \textit{Right:} the near infrared CMD of \tuc{}, transformed into the initial mass -- oxygen abundance space using the calculated isochrones. The transformed \textit{red tail}, \textit{ridgeline} and \textit{blue tail} isochrones are indicated with the \textit{red}, \textit{black} and \textit{blue} vertical lines, respectively. \textit{Bottom:} combined \ce{[O/Fe]} distribution at $\qty{0.15}{\Msun}<M<\qty{0.4}{\Msun}$, overplotted on the distribution of spectroscopic measurements in the literature (Section~\ref{sec:nominal_chem}). The measurements of mass and oxygen abundance shown in this figure are available as a machine-readable table in the digital version of this publication.}
    \label{fig:oxygen_by_mass}
\end{figure}

To obtain the desired transformation, we first generated a $1000\times300$ regular grid of synthetic stars at $1000$ evenly spaced initial masses from the minimum ($\approx\qty{0.045}{\Msun}$) to the maximum ($\approx\qty{0.875}{\Msun}$) value within the range of all three isochrones, and $300$ evenly spaced values of \ce{[O/Fe]} from $\qty{-0.57}{dex}$ to $\qty{0.78}{dex}$. The chosen range of oxygen abundances corresponds to tripled differences between the \textit{ridgeline} oxygen abundance and the other two isochrones. Each synthetic star was projected onto the near infrared color-magnitude plane by, first, linearly interpolating the CMD of each of the three isochrones to the mass of the synthetic star and, second, linearly interpolating and extrapolating the resulting three points to the oxygen abundance of the synthetic star. The curvilinear transformation of the observed CMD of the cluster was carried out by identifying the closest synthetic star to each real observation in the color-magnitude space, and assigning the mass and \ce{[O/Fe]} values of the closest synthetic star to the real star. This method is unreliable above the main sequence \review{knee}, where all three isochrones merge and the curvilinear transformation becomes non-unique. We carry out this analysis below the \review{knee} ($\qty{0.1}{\Msun}\leq M\leq \qty{0.5}{\Msun}$), where the isochrones are well-separated, and the transformation is accurate to $\qty{e-3}{\Msun}$ in mass and $\qty{e-2}{dex}$ in oxygen abundance, as estimated by applying the transformation to $\num{e4}$ additional synthetic stars with randomly chosen masses and oxygen abundances. We further restrict our analysis to stars that satisfy $\qty{-0.27}{dex}<\ce{[O/Fe]}<\qty{0.63}{dex}$ (doubled differences between the calculated isochrones) to avoid excessive extrapolation and to discard the occasional outlying measurements that fell beyond the region of the CMD covered with synthetic stars.

\begin{deluxetable}{lrlr}
\tablecaption{Comparison of synthetic and real distributions of spectroscopically inferred \ce{[O/Fe]} \label{table:spect_vs_phot}}
\tablewidth{0pt}
\tablehead{
\colhead{Statistic} & \multicolumn{2}{c}{Synthetic}  & \colhead{Real}
}
\startdata
Mean   &    $0.232$ &    $\pm0.024$  &    $0.174$  \\
Standard deviation   &    $0.253$ &    $\pm0.015$  &    $0.214$  \\
5\textsuperscript{th} percentile   &    $-0.198$ &    $\pm0.047$  &    $-0.170$  \\
25\textsuperscript{th} percentile   &    $0.058$ &    $\pm0.036$  &    $0.030$  \\
75\textsuperscript{th} percentile   &    $0.414$ &    $\pm0.029$  &    $0.320$  \\
95\textsuperscript{th} percentile   &    $0.616$ &    $\pm0.038$  &    $0.476$  \\
\hline
\enddata
\tablecomments{All values are quoted in \qty{}{dex}. The indicated uncertainty of the synthetic values was calculated as the standard deviation across $10^5$ Monte-Carlo trials.}
\end{deluxetable}

The result of the transformation is shown in the \textit{right} panel of Figure~\ref{fig:oxygen_by_mass}. The distribution of \ce{[O/Fe]} in selected mass bins is shown in the left panel of the figure. We note that the highest mass bin ($\qty{0.4}{\Msun}<M<\qty{0.5}{\Msun}$) is an unreliable indicator of the oxygen distribution due to the proximity of the main sequence \review{knee}, where the isochrone fit is noticeably poorer. The lower half of the lowest mass bin ($\qty{0.1}{\Msun}<M<\qty{0.15}{\Msun}$) is visibly affected by the contamination of the CMD by SMC members. The oxygen abundances within the remaining mass range ($\qty{0.15}{\Msun}<M<\qty{0.4}{\Msun}$) were combined into a single oxygen abundance distribution in the lower panel of Figure~\ref{fig:oxygen_by_mass}. A histogram of the spectroscopic measurements of higher-mass stars ($M\gtrsim\qty{0.9}{\Msun}$) is plotted in the same panel for reference.

To compare the spectroscopic and photometric abundances quantitatively, we began with a null-hypothesis that (1) the photometric distribution is free of systematic errors and represents the true chemistry in the lower main sequence, and (2) photometric and spectroscopic abundances are identical, i.e. there is no dependence of \ce{[O/Fe]} on stellar mass. The spectroscopic distribution of oxygen, as described in Section~\ref{sec:nominal_chem}, consists of $117$ individual measurements with published uncertainties. We generated $10^5$ synthetic equivalents of the spectroscopic dataset by drawing $117$ random measurements from the photometric distribution for each trial and applying simulated Gaussian noise according to the published spectroscopic uncertainties. For each synthetic set, we estimated the mean, standard deviation, 5\textsuperscript{th}, 25\textsuperscript{th}, 75\textsuperscript{th} and 95\textsuperscript{th} percentiles. The averages and spreads in those statistics across all $10^5$ Monte-Carlo trials are summarized in Table~\ref{table:spect_vs_phot}, alongside the equivalent values calculated for the real spectroscopic dataset. We note that the 25\textsuperscript{th} and 75\textsuperscript{th} percentiles fall within the range of \ce{[O/Fe]} covered by the model isochrones and are, therefore, immune to extrapolation errors. All other statistics listed in the table are sensitive to the systematic offsets introduced by extrapolation.

The null-hypothesis was rejected, since some of the statistical parameters of the real spectroscopic distribution fall outside the corresponding synthetic ranges by over $\sim3$ standard deviations, most notably including the extrapolation-immune 75\textsuperscript{th} percentile. If the systematic errors in the photometrically inferred abundances are the dominant contributor to the estimated discrepancy, their magnitude appears to vary from $\approx\qty{0.03}{dex}$ at the oxygen-poor tail of the distribution (inferred from the 5\textsuperscript{th} percentile) to $\approx\qty{0.15}{dex}$ at the oxygen-rich tail (inferred from the 95\textsuperscript{th} percentile). We note that the latter value is comparable to the average spectroscopic uncertainty of $\qty{0.147}{dex}$ across the $117$ measurements.

Alternatively, a genuine variation in chemistry with stellar mass may be responsible for the discrepancy between spectroscopic and photometric values. Under the standard assumption that the enriched population of \tuc{} is oxygen-deficient \citep{CNO_sum_constant}, the values of the higher percentiles are expected to be primarily determined by the oxygen content of the primordial population, while the values of the lower percentiles would be mostly set by \ce{[O/Fe]} of the enriched population. As discussed in Section~\ref{sec:introduction}, the chemistry of the enriched population may depend on stellar mass in concurrent formation models. We, however, disfavor this explanation, since the spectroscopic/photometric discrepancy in Table~\ref{table:spect_vs_phot} appears largest at the oxygen-rich (primordial) tail of the \ce{[O/Fe]} distribution, rather than the anticipated oxygen-poor (enriched) tail.

Both photometric and spectroscopic distributions in the lower panel of Figure~\ref{fig:oxygen_by_mass} exhibit a clear negative skewness (i.e. the oxygen-poor tail is longer than the oxygen-rich one). To assess the statistical significance of this feature, we calculated the Fisher-Pearson coefficient of skewness for both distributions, obtaining $-0.21\pm0.03$ and $-0.43\pm0.19$ for the photometric and spectroscopic cases respectively. Hence, spectroscopic and photometric skewness estimates are consistent with each other, and indicate that the distribution of \ce{[O/Fe]} in \tuc{} is indeed negatively skewed. While both skewness estimates have high confidence, the photometric value is vastly more statistically significant than its spectroscopic counterpart ($8.4$ vs $\sim2.2$ standard deviations, respectively), due to the number of photometric measurements ($4862$) exceeding the number of spectroscopic measurements ($117$) by over an order of magnitude.
    
    \subsection{Mass function} \label{sec:mass_function}
    The mass functions of globular clusters retain a footprint of their dynamical evolution \citep{Sollima_1,Sollima_2} and may vary between populations if they have distinct kinematic properties. Furthermore, an estimate of the mass function is required to predict the luminosity function and the CMD of the substellar sequence. In this sub-section, we seek a mass function that can reproduce the observed \texttt{F160W} magnitude distribution of \tuc{}, assuming that the isochrones calculated earlier and the oxygen abundance distribution derived above accurately capture the properties of the cluster. The choice of the photometric band is motivated by the fact that the oxygen abundances were inferred from the near infrared CMD and the fact that lower main sequence members are brighter in \texttt{F160W} than in \texttt{F110W} (Figure~\ref{fig:best_fit}). We also assumed that the mass function ($\xi$) is of the form of a broken power law \citep{Kroupa}:

\begin{equation}
    \xi(M)\propto\begin{cases}
        M^{-\beta},& \text{if } M>M_\mathrm{bp}\\
        M^{-\gamma},& \text{if } M\leq M_\mathrm{bp}\\
    \end{cases}
    \label{eq:powerlaw}
\end{equation}

Here, $\beta$ and $\gamma$ are the power law indices in the high- and low-mass regimes respectively, separated by the break-point stellar mass $M_\mathrm{bp}$. We keep all three parameters as free variables. The magnitude function, $\phi(m)$, where $m$ is the apparent magnitude of the star (in our case, \texttt{F160W}), is related to the mass function as in Equation~\ref{eq:MF_to_LF}:

\begin{equation}
    \phi(m)=-\xi\left(M(m)\right) \frac{dM(m)}{dm}
    \label{eq:MF_to_LF}
\end{equation}

In Equation~\ref{eq:MF_to_LF}, we treat the stellar mass, $M$, as a function of magnitude, $m$, as given by the mass-magnitude relationship provided by the isochrone. In practice, the mass-magnitude relationship is sampled at a set of masses, $M_i$, and magnitudes, $m_i$, corresponding to the initial masses of the calculated evolutionary models. We therefore rewrite Equation~\ref{eq:MF_to_LF} in the form of finite differences:

\begin{equation}
    \phi\left(\frac{m_i+m_{i+1}}{2}\right)=-\xi\left(\frac{M_i+M_{i+1}}{2}\right) \frac{M_{i+1}-M_i}{m_{i+1}-m_i}
    \label{eq:MF_to_LF_FD}
\end{equation}

\begin{figure}[ht!]
    \centering
    \includegraphics[width=1\columnwidth]{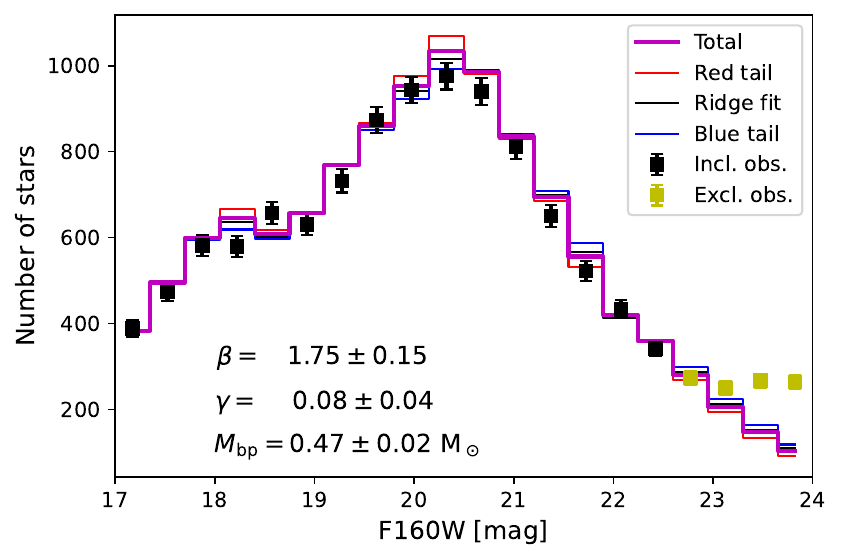}
    \caption{The observed magnitude function of \tuc{} in \texttt{F160W} with the best-fit theoretical magnitude function in \textit{magenta}. Additional magnitude functions that use only one of the theoretical isochrones instead of the abundance-weighted average are shown for comparison. The displayed error bars are Poisson counting errors. The observed counts in the four faintest bins are excluded from this analysis due to contamination by SMC members.}
    \label{fig:LF_fit}
\end{figure}

Here, the magnitude function is evaluated at the mid-points between the adjacent evolutionary models on the mass grid. We calculated three distinct magnitude functions for each of the three final isochrones (\textit{red tail}, \textit{blue tail} and \textit{ridgeline}). The magnitude function for an arbitrary \ce{[O/Fe]} can be obtained by linearly interpolating/extrapolating between the three calculated magnitude functions to the desired oxygen abundance. The combined magnitude function for the entire cluster was computed as the average of individual magnitude functions, evaluated for the oxygen abundance of every star that satisfies $\qty{0.15}{\Msun}<M<\qty{0.4}{\Msun}$ and $\qty{-0.27}{dex}<\ce{[O/Fe]}<\qty{0.63}{dex}$ (the same restrictions as in the lower panel of Figure~\ref{fig:oxygen_by_mass}). Finally, the magnitude function was binned into $20$ uniform bins in the range $17<m<24$ using trapezoid integration. The four faintest bins with bin centers at \texttt{F160W} $>22.5$ were excluded from our analysis due to contamination by SMC members and potential incompleteness of the photometric sample.

To determine the parameters of the mass function, $\beta$, $\gamma$ and $M_\mathrm{bp}$, we applied the same binning to the observed magnitude function in \texttt{F160W} and fitted the theoretical magnitude function to the result, using least squares regression, weighted by the Poisson errors in each bin. The observed magnitude function and the theoretical best fit are shown in Figure~\ref{fig:LF_fit}. The best-fit parameters of the mass function are $\beta=1.75\pm0.15$, $\gamma=0.08\pm0.04$ and $M_\mathrm{bp}=\qty{0.47\pm0.02}{\Msun}$. In addition to the combined magnitude function, three more theoretical magnitude functions were calculated for the same parameters as the best fit, but using only one of the three isochrones instead of a weighted average of all three. As seen in Figure~\ref{fig:LF_fit}, the scatter among the theoretical magnitude functions is consistent with the Poisson counting errors in the corresponding magnitude bins, suggesting that for main sequence members the distribution of oxygen abundance cannot be inferred from the observed luminosity function of the cluster.
    
    \subsection{Brown dwarfs in \tuc{}} \label{sec:brown_dwarfs}
    \begin{figure}[ht!]
    \centering
%    \begin{interactive}{animation}{figures/fig_9_animated.mp4}
    \includegraphics[width=1\columnwidth]{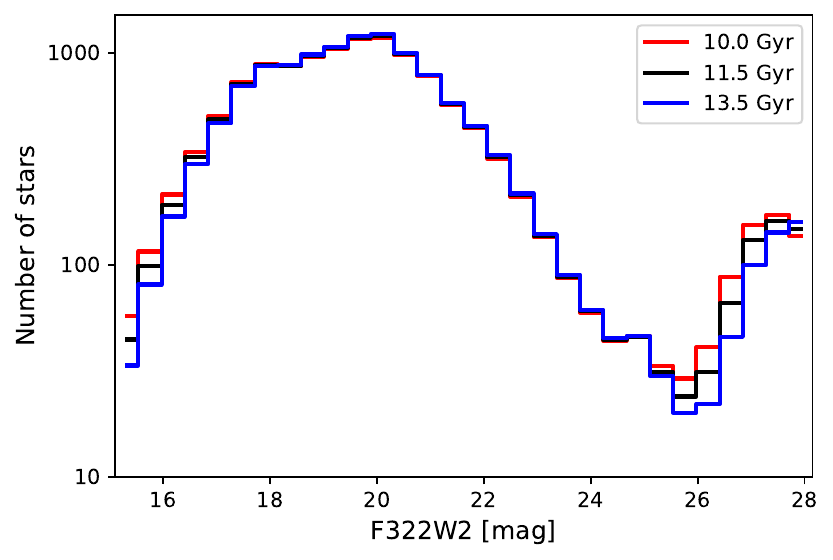}
%    \end{interactive}
    \caption{Predicted magnitude distribution of stars and brown dwarfs in \tuc{} at three different ages. The bright and faint peaks represent the main sequence and the brown dwarfs of the cluster, respectively, with a clear stellar/substellar gap in between. The distribution is normalized to the number of stars in the dataset (Section~\ref{sec:data}). \review{An animated version of this figure is available in the digital version of the publication. The animation lasts $6$ seconds and includes $30$ frames. The animation shows the evolution of the magnitude function from $\qty{0.1}{Gyr}$ to $\qty{13.5}{Gyr}$. Initially, the magnitude function only has one peak around $m\sim 20$ corresponding to the main sequence of the cluster. The stellar/substellar gap forms at $m\sim23.5$ around $\qty{0.5}{Gyr}$. Over time, it gradually deepens, widens, and shifts to fainter magnitudes.}}
    \label{fig:LF_predicted}
\end{figure}

\begin{figure}[ht!]
    \centering
    \includegraphics[width=1\columnwidth]{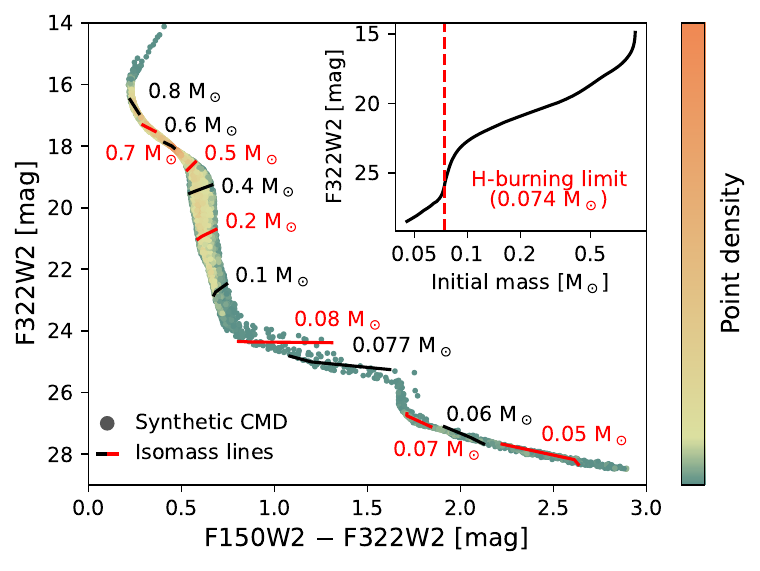}
    \caption{Predicted CMD of \tuc{} from the subgiant branch to the substellar sequence. Isomass lines are shown for selected masses. The color scatter in the figure is derived from the inferred distribution of \ce{[O/Fe]} near the end of the main sequence. The inset sub-figure in the upper right corner shows the mass-magnitude relationship for the \textit{ridgeline} isochrone with the hydrogen-burning limit highlighted. The isochrones used to construct the CMD in this figure are available as machine-readable tables in the digital version of this publication.}
    \label{fig:CMD_predicted}
\end{figure}

In this sub-section, we predict the colors, magnitudes and number densities of the brown dwarfs in \tuc{}, as they may be observed with \textit{JWST} \textit{NIRCam}. The \textit{JWST} color-magnitude space considered in this study is \texttt{F322W2} vs \texttt{F150W2} -- \texttt{F322W2}, since brown dwarfs are most likely to be detected in wide bands at long wavelengths due to their faint magnitudes and red colors. Our predictions are based on the assumption that the mass function (Equation~\ref{eq:powerlaw} with the best-fit parameters in Figure~\ref{fig:LF_fit}), the calculated isochrones (Figure~\ref{fig:best_fit}) and the inferred distribution of chemical abundances (Figure~\ref{fig:oxygen_by_mass}) remain unchanged in the substellar regime.

We re-calculated the synthetic magnitude function of \tuc{} in $30$ evenly spaced magnitude bins within range of all three theoretical isochrones, using the best-fit mass function parameters. The calculations were carried out for a variety of ages between $\qty{0.1}{Gyr}$ and $\qty{13.5}{Gyr}$, including $\qty{11.5}{Gyr}$ (best-fit age). The requirement to vary the age of the cluster for this part of the analysis necessitated the use of interpolated bolometric corrections for synthetic photometry instead of atmosphere hammering. The resulting magnitude functions for three of the considered ages are shown in \review{the static preview} of Figure~\ref{fig:LF_predicted}. At those ages, the magnitude function exhibits a clear stellar/substellar gap at $m\sim 26$, where the brown dwarf density per magnitude is reduced by nearly an order of magnitude (compared to the maximum density within the modelled range that is predicted to occur around $m=27.5$). While the lower main sequence is virtually unaffected by the cluster age, both the turn-off point and the substellar sequence undergo noticeable evolution. \review{The equivalent magnitude functions for all other ages are available as an animated figure in the digital version of this paper.}

A synthetic CMD based on the inferred oxygen abundance in Figure~\ref{fig:oxygen_by_mass}, the isochrones in Figure~\ref{fig:best_fit} and the predicted mass function at $\qty{11.5}{Gyr}$ is shown in Figure~\ref{fig:CMD_predicted}. Lines of constant mass (isomass lines) for selected masses and $\qty{-0.27}{dex}<\ce{[O/Fe]}<\qty{0.63}{dex}$ are also shown for reference. Since the isochrones of the substellar sequence in \tuc{} appear almost exactly parallel to the isomass lines, the effect of evolution on the brown dwarf colors is highly degenerate with \ce{[O/Fe]}. This presents a potential challenge to the use of brown dwarfs as chemical tracers, since the masses of individual brown dwarfs (and, hence, their evolutionary phases) are not \textit{a priori} known. The degeneracy also reduces the observed photometric scatter, making the substellar sequence far narrower than the main sequence in the CMD space.

The predicted mass-magnitude relationship for the \textit{ridgeline} isochrone is shown in an inset sub-figure of Figure~\ref{fig:CMD_predicted} with the hydrogen-burning limit (HBL) highlighted. We define the HBL as the mass, at which the energy output from nuclear reactions contributes $50\%$ of the stellar luminosity at $\qty{11.5}{Gyr}$. In this case, the HBL was computed as $\qty{0.074}{\Msun}$.

\section{Discussion and conclusion} \label{sec:conclusion}
We developed a general method for determining the chemical compositions, ages and mass functions of globular clusters from the observed CMDs in optical and near infrared bands. Our method relies on state-of-the-art model atmospheres and evolutionary models that are fully self-consistent and incorporate the full set of non-solar abundances in every component, including the interior structure and evolution, nuclear processing, atmosphere-interior coupling, atmospheric structures and spectral synthesis. Our modelling framework was applied to the brightest mono-metallic globular cluster \tuc{}. We reproduce for the first time the observed scatter in the photometric colors of main sequence stars without any \textit{a priori} assumptions of its magnitude. We also provide the first measurements of chemical compositions for individual stars of the lower main sequence in a globular cluster (Figure~\ref{fig:oxygen_by_mass}). An extension of our models to the substellar regime predicts the expected colors and magnitudes of brown dwarfs in the cluster, reproducing the anticipated stellar/substellar gap. The predicted brown dwarf CMD for the first time incorporates the inferred distribution of chemical abundances. The best-fit parameters of \tuc{} determined in this study are listed in Table~\ref{table:best_fit}. The key findings are as follows:

\begin{itemize}
    \item The photometric scatter of \tuc{} in both optical and near infrared bands can be reproduced with unaltered average spectroscopic abundances for all elements (Appendix~\ref{sec:nominal_chem_table}) with the exception of \ce{[O/Fe]} and \ce{[Ti/Fe]}. Constraining the abundances of elements other than oxygen and titanium from the CMDs considered in this study is challenging due to the subdominant effect of those elements on stellar atmospheres, as illustrated in Figure~\ref{fig:variations}.
    \item The observed photometric scatter is predominantly driven by \ce{[O/Fe]}. Our best-fit model CMD is consistent with the absence of star-to-star variations in \ce{[Ti/Fe]}, in agreement with the spectroscopic measurements and the mono-metallic nature of \tuc{}.
    \item The photometric distribution of the oxygen abundance estimated in this study (Figure~\ref{fig:oxygen_by_mass}) at $M<\qty{0.4}{\Msun}$ is statistically consistent with the spectroscopic distribution, inferred from evolved stars, under the assumption of systematic errors in the photometric estimates of order $\qty{0.15}{dex}$. These errors are comparable to the experimental uncertainties in available spectroscopic measurements, demonstrating that lower main sequence CMDs of globular clusters may be used in chemical analyses.
    \item The discrepancy between the spectroscopic and photometric \ce{[O/Fe]} values appears largest at the oxygen-rich (\review{primordial}) tail of the distribution. If a genuine mass dependence of chemistry makes a significant contribution to the observed discrepancy in addition to the suspected systematic offset, the variation of \ce{[O/Fe]} with stellar mass would be most prominent in the \review{enriched} population of the cluster instead of the \review{primordial} population. Since this conclusion contradicts the theoretical expectation, we disfavor this explanation and suggest that the mass dependence of chemistry makes a subdominant contribution compared to the systematic offset, if any. We therefore expect the difference in \ce{[O/Fe]} between the lower main sequence and evolved stars to be less than the estimated systematic offset of $~\qty{0.15}{dex}$.
    \item Corroborating our previous study of \omegacen{}, we confirm that the \textit{JWST} CMD of \tuc{} is expected to have a stellar/substellar gap below the end of the main sequence, followed by a large number of brown dwarfs (Figure~\ref{fig:LF_predicted}). The gap occurs around \texttt{F322W2}$\sim \qty{26}{mag}$ and the maximum density of brown dwarfs within the modelling range is attained at \texttt{F322W2}$\sim \qty{27.5}{mag}$. Our evolutionary models suggest that stellar/substellar gaps form around the age of $\qty{0.5}{Gyr}$ in globular clusters similar to \tuc{}, and get deeper and wider over time.
    \item While our models indicate that the colors of brown dwarfs are highly sensitive to chemical abundances (Figure~\ref{fig:variations}), the apparent degeneracy between their isomass lines and isochrones (Figure~\ref{fig:CMD_predicted}) will make the inference of oxygen abundance challenging in the considered color-magnitude space (\texttt{F322W2} vs \texttt{F150W2} -- \texttt{F322W2}). On the other hand, the brown dwarf magnitude function will be sensitive to chemistry, introducing additional systematic biases to age estimates inferred from the stellar/substellar gap.
\end{itemize}

The constraints on the mass dependence of the oxygen abundance inferred in this study restrict the allowed parameter space of concurrent formation MP models, in agreement with previous studies of the photometric spread near the end of the main sequence \citep{JWST_low_MS_phot_1,JWST_low_MS_phot_2}. However, a detailed model of pollution as a function of stellar mass is required to carry out a thorough evaluation of proposed concurrent formation models, which may be challenging to derive due to the uncertain properties and evolution of circumstellar disks.

Future observations of the substellar sequences in nearby globular clusters, including \tuc{}, will extend the mass baseline of photometric analysis by nearly an order of magnitude and explore the abundances of other elements such as carbon and nitrogen that are far more important in the atmospheres of brown dwarfs than main sequence stars (Figure~\ref{fig:variations}). The aforementioned isomass-isochrone degeneracy suggests that the luminosity function of brown dwarfs in globular clusters needs to be considered in conjunction with the CMD when deriving photometric chemical abundances. We further emphasize that \textit{JWST} and \textit{HST} bands other than the ones considered here may be more sensitive to chemical variations in substellar atmospheres, necessitating a follow-up study that explores all possible color combinations rather than a small subset of filters used in this paper.

We note that the predicted substellar CMD of \tuc{} in Figure~\ref{fig:CMD_predicted} was calculated under the assumption that the variation in \ce{[O/Fe]} alone is sufficient to reproduce the photometric scatter. While this assumption was shown to be accurate for the main sequence stars, it likely has limited validity in the substellar regime and a more accurate theoretical CMD may be obtained by calculating additional isochrones that incorporate the spectroscopically inferred scatter in other elements.

If the spectroscopic/photometric discrepancy discussed here and in Section~\ref{sec:abundances} is primarily caused by systematic offsets, their nature must be investigated in a future study. In particular, the accuracy of interpolation and extrapolation of best-fit theoretical isochrones can be verified by a denser sampling of \ce{[O/Fe]} with calculated stellar models. The accuracy of the models themselves can be assessed by comparing synthetic spectra to spectroscopic observations of nearby metal-poor stars and brown dwarfs.

\pagebreak
%\begin{acknowledgements}
    We acknowledge the funding support from Hubble Space Telescope (HST) Program GO-15096, provided by NASA through a grant from the Space Telescope Science Institute, which is operated by the Association of Universities for Research in Astronomy, Incorporated, under NASA contract NAS5-26555. The computational demand of this study was met by the Extreme Science and Engineering Discovery Environment (XSEDE), supported by NSF grant ACI-1548562. Some of the software used in this study was produced with assistance from ChatGPT 3.5. ChatGPT is a large language model developed and maintained by OpenAI. This work was conducted at University of California San Diego, which was built on the unceded territory of the Kumeyaay Nation. Today, the Kumeyaay people continue to maintain their political sovereignty and cultural traditions as vital members of the San Diego community.  We acknowledge their tremendous contributions to our region and thank them for their stewardship.
%\end{acknowledgements}

% \software{
% \texttt{Astropy} \citep{astropy_1,astropy_2},  
% \texttt{Matplotlib} \citep{matplotlib},
% \texttt{NumPy} \citep{numpy}, 
% \texttt{SciPy} \citep{scipy}
% }

% \facilities{{\em HST} (\textit{ACS}, \textit{WFC3})}

\clearpage

\appendix

\section{Nominal composition} \label{sec:nominal_chem_table}
Table~\ref{tab:nominal_chem} lists the nominal composition of \tuc{} adopted in this work, based on spectroscopic observations of giant and sub-giant members of the cluster in the literature. The derivation of the abundances listed in the table is described in Section~\ref{sec:nominal_chem}. The physical member-to-member spread provided in the table ($s^{(X)}$) was calculated using Equation~\ref{eq:abundance_spread}.

\startlongtable
\begin{deluxetable}{l|ccccc}
\tablecaption{Average chemical composition of \tuc{} inferred from spectroscopic measurements in literature\label{tab:nominal_chem}}
\tablewidth{\textwidth}
\tablehead{
\colhead{} & \colhead{Best estimate} & \colhead{Uncertainty} & \colhead{Physical spread} & \colhead{\# of measurements} & \colhead{Reference(s)}
}
\startdata
\ce{[Fe/H]}   &   $-0.75$   &   $ 0.01$   &   $-$ & $281$ & $(1)(2)(3)$ \\
\ce{[Al/Fe]}   &   $ 0.35$   &   $ 0.01$   &   $ 0.06$ & $195$ & $(1)(2)(3)$ \\
\ce{[Ba/Fe]}   &   $ 0.25$   &   $ 0.07$   &   $-$ & $13$ & $(1)$ \\
\ce{[C /Fe]}   &   $-0.25$   &   $ 0.01$   &   $ 0.10$ & $70$ & $(3)$ \\
\ce{[Ca/Fe]}   &   $ 0.26$   &   $ 0.01$   &   $-$ & $162$ & $(1)(2)$ \\
\ce{[Ce/Fe]}   &   $-0.04$   &   $ 0.10$   &   $-$ & $11$ & $(1)$ \\
\ce{[Co/Fe]}   &   $-0.00$   &   $ 0.02$   &   $-$ & $13$ & $(1)$ \\
\ce{[Cr/Fe]}   &   $-0.04$   &   $ 0.03$   &   $-$ & $13$ & $(1)$ \\
\ce{[Cu/Fe]}   &   $-0.14$   &   $ 0.10$   &   $-$ & $13$ & $(1)$ \\
\ce{[Dy/Fe]}   &   $ 0.70$   &   $ 0.07$   &   $-$ & $13$ & $(1)$ \\
\ce{[Eu/Fe]}   &   $ 0.44$   &   $ 0.01$   &   $-$ & $150$ & $(1)(2)$ \\
\ce{[La/Fe]}   &   $ 0.20$   &   $ 0.01$   &   $-$ & $144$ & $(1)(2)$ \\
\ce{[Mg/Fe]}   &   $ 0.35$   &   $ 0.01$   &   $-$ & $87$ & $(1)(3)$ \\
\ce{[Mn/Fe]}   &   $-0.19$   &   $ 0.04$   &   $-$ & $13$ & $(1)$ \\
\ce{[Mo/Fe]}   &   $ 0.55$   &   $ 0.04$   &   $-$ & $13$ & $(1)$ \\
\ce{[N /Fe]}   &   $ 0.85$   &   $ 0.05$   &   $ 0.33$ & $54$ & $(3)$ \\
\ce{[Na/Fe]}   &   $ 0.27$   &   $ 0.01$   &   $ 0.15$ & $236$ & $(1)(2)(3)$ \\
\ce{[Nd/Fe]}   &   $ 0.04$   &   $ 0.07$   &   $-$ & $13$ & $(1)$ \\
\ce{[Ni/Fe]}   &   $-0.06$   &   $ 0.01$   &   $-$ & $174$ & $(1)(2)(3)$ \\
\ce{[O /Fe]}   &   $ 0.18$   &   $ 0.02$   &   $ 0.16$ & $117$ & $(1)(2)$ \\
\ce{[Pr/Fe]}   &   $-0.03$   &   $ 0.06$   &   $-$ & $13$ & $(1)$ \\
\ce{[Ru/Fe]}   &   $ 0.50$   &   $ 0.04$   &   $-$ & $13$ & $(1)$ \\
\ce{[Sc/Fe]}   &   $ 0.21$   &   $ 0.05$   &   $-$ & $13$ & $(1)$ \\
\ce{[Si/Fe]}   &   $ 0.32$   &   $ 0.01$   &   $-$ & $222$ & $(1)(2)(3)$ \\
\ce{[Ti/Fe]}   &   $ 0.34$   &   $ 0.01$   &   $-$ & $163$ & $(1)(2)$ \\
\ce{[V /Fe]}   &   $ 0.17$   &   $ 0.04$   &   $-$ & $13$ & $(1)$ \\
\ce{[Y /Fe]}   &   $ 0.07$   &   $ 0.05$   &   $-$ & $13$ & $(1)$ \\
\ce{[Zn/Fe]}   &   $ 0.26$   &   $ 0.04$   &   $-$ & $13$ & $(1)$ \\
\ce{[Zr/Fe]}   &   $ 0.41$   &   $ 0.06$   &   $ 0.10$ & $13$ & $(1)$ \\
\enddata

\tablecomments{
All values are quoted in \qty{}{dex} with respect to solar abundances.
(1) -- \citet{nominal_T14}. (2) -- \citet{nominal_C14}. (3) -- \citet{nominal_M16}.}

\end{deluxetable}

\bibliography{references}{}
\bibliographystyle{aasjournal}

\end{document}